\documentclass[prd,aps,floats,floatfix,eqsecnum,nofootinbib]{revtex4}
\usepackage{amsmath,amssymb,verbatim,epsfig,graphicx,rotating}
\usepackage{psfrag}
\newcommand{\be}{\begin{equation}}
\newcommand{\ee}{\end{equation}}
\newcommand{\bea}{\begin{eqnarray}}
\newcommand{\eea}{\end{eqnarray}}
\begin{document}
\title{MCMC analysis of WMAP3 and SDSS data points to broken symmetry inflaton potentials
and provides a lower bound on the tensor to scalar ratio.}
\author{\bf C. Destri$^{(a)}$} \email{Claudio.Destri@mib.infn.it}
\author{\bf H. J. de Vega$^{(b,c)}$} \email{devega@lpthe.jussieu.fr}
\author{\bf N. G. Sanchez$^{(c)}$} \email{Norma.Sanchez@obspm.fr}
\affiliation{$^{(a)}$ Dipartimento di Fisica G. Occhialini, Universit\`a
Milano-Bicocca Piazza della Scienza 3, 20126 Milano and
INFN, sezione di Milano, via Celoria 16, Milano Italia\\
$^{(b)}$ LPTHE, Laboratoire Associ\'e au CNRS UMR 7589,\\
Universit\'e Pierre et Marie Curie (Paris VI) et Denis Diderot (Paris VII),\\
Tour 24, 5 \`eme. \'etage, 4, Place Jussieu, 75252 Paris, Cedex 05, France.\\ 
$^{(c)}$ Observatoire de Paris, LERMA, Laboratoire Associ\'e au CNRS UMR 8112,
 \\61, Avenue de l'Observatoire, 75014 Paris, France.}
\begin{abstract}
We perform a MCMC (Monte Carlo Markov  Chains) analysis of the available 
CMB and LSS data (including the three years WMAP data)
with single field slow-roll new inflation and chaotic inflation models.
We do this within our approach to inflation as an
effective field theory in the Ginsburg-Landau spirit
with fourth degree trinomial potentials in the  inflaton field  $ \phi $.
We derive explicit formulae and study in detail
the spectral index $ n_s $ of the adiabatic fluctuations, the ratio $ r $
of tensor to scalar fluctuations and the running  index $ d n_s/d
\ln k $. We use these analytic formulas as hard constraints on
$ n_s $ and $ r $ in the MCMC analysis.  
Our analysis differs in this {\bf crucial} aspect from previous MCMC studies in the literature
involving the WMAP3 data. Our results are as follow:
(i) The data strongly indicate the {\bf breaking} (whether spontaneous or explicit)
of the $ \phi \to - \phi $ symmetry of the inflaton potentials both for new and
for chaotic inflation. (ii) Trinomial new inflation naturally satisfies this requirement 
and provides an excellent fit to the data. (iii) Trinomial chaotic inflation produces 
the best fit in a very narrow corner of the parameter space. (iv) The chaotic symmetric 
trinomial potential is almost certainly {\bf ruled out} (at $95\%$CL).
In trinomial chaotic inflation the MCMC runs go towards a potential in the
{\em boundary} of the parameter space and which ressembles a spontaneously symmetry broken
potential of new inflation.
(v) The above results and further physical analysis here lead us to conclude that {\bf new
inflation} gives the best description of the data. (vi) We find a lower bound for $ r $
within trinomial new inflation potentials: $ r > 0.016 \;  (95\% \; {\rm CL}) $ and $ r >
0.049 \;  (68\% \;   {\rm CL}) $.  (vii) The preferred new inflation trinomial potential 
is a double well, even function of the field with a moderate quartic coupling yielding 
as most probable values: $ n_s \simeq 0.958 ,\; r\simeq 0.055 $. This value for $ r $ is
within reach of forthcoming CMB observations.
\end{abstract} 

\date{\today}
\maketitle
\tableofcontents

\section{Introduction and Results}

Inflation was introduced to solve several outstanding problems of the
standard Big Bang model \cite{guth} and has now become an important part of
the standard cosmology. At the same time, it provides a natural mechanism
for the generation of scalar density fluctuations that seed large scale
structure, thus explaining the origin of the temperature anisotropies in
the cosmic microwave background (CMB), as well as that of tensor
perturbations (primordial gravitational waves) \cite{mukyotr,libros,hu}.

\medskip

A distinct aspect of inflationary perturbations is that these are generated
by quantum fluctuations of the scalar field(s) that drive inflation. After
their wavelength becomes larger than the Hubble radius, these fluctuations
are amplified and grow, becoming classical and decoupling from causal
microphysical processes. Upon re-entering the horizon, during the matter
era, these classical perturbations seed the inhomogeneities which generate
structure upon gravitational collapse \cite{mukyotr,libros}. A great
diversity of inflationary models predict fairly generic features: a
gaussian, nearly scale invariant spectrum of (mostly) adiabatic scalar and
tensor primordial fluctuations, making the inflationary paradigm fairly
robust. The gaussian, adiabatic and nearly scale invariant spectrum of
primordial fluctuations provide an excellent fit to the highly precise
wealth of data provided by the Wilkinson Microwave Anisotropy Probe
(WMAP)\cite{WMAP,WMAP3}.  Perhaps the most striking validation of inflation
as a mechanism for generating \emph{superhorizon} (`acausal') fluctuations
is the anticorrelation peak in the temperature-polarization (TE) angular
power spectrum at $l \sim 150$ corresponding to superhorizon
scales \cite{WMAP}. The confirmation of many of the robust predictions of
inflation by current high precision observations places inflationary
cosmology on solid grounds.

\medskip

Amongst the wide variety of inflationary scenarios, single field slow roll
models provide an appealing, simple and fairly generic description of
inflation. Its simplest implementation is based on a scalar field (the
inflaton) whose homogeneous expectation value drives the dynamics of the
scale factor, plus small quantum fluctuations. The inflaton potential is
fairly flat during inflation. This flatness not only leads to a slowly
varying Hubble parameter, hence ensuring a sufficient number of e-folds,
but also provides an explanation for the gaussianity of the fluctuations as
well as for the (almost) scale invariance of their power spectrum. A flat
potential precludes large non-linearities in the dynamics of the
\emph{fluctuations} of the scalar field.

\medskip

The current WMAP data seem to validate the simpler one-field slow roll
scenario \cite{WMAP,WMAP3}. Furthermore, because the potential is flat the
scalar field is almost {\bf massless}, and modes cross the horizon with an
amplitude proportional to the Hubble parameter. This fact combined with a
slowly varying Hubble parameter yields an almost scale invariant primordial
power spectrum.  The slow-roll approximation has been recently cast as a $
1/N $ expansion \cite{1sN}, where $ N \sim 50 $ is the number of efolds
before the end of inflation when modes of cosmological relevance today
first crossed the Hubble radius.

\medskip

The observational progress permits to start to discriminate among different
inflationary models, placing stringent constraints on them. The upper bound
on the ratio $ r $ of tensor to scalar fluctuations obtained by WMAP
\cite{WMAP,WMAP3} {\bf necessarily} implies the presence of a {\bf mass
  term} in the single field inflaton potential and therefore rules out the
massless monomial $ \phi^4 $ potential \cite{1sN,WMAP3}. Hence, as minimal single
field model, one should consider a sufficiently general quartic polynomial,
that is the trinomial potential.

Besides its simplicity, the trinomial potential is a physically well
motivated potential for inflation in the context
of the Ginsburg-Landau approach to effective field theories (see for example ref.\cite{quir}).
This potential is rich enough to describe the physics of inflation and
accurately reproduce the WMAP data \cite{WMAP,WMAP3}.

The slow-roll expansion plus the WMAP data constraints the inflaton potential
to have the form \cite{1sN}
\be \label{Vi} 
V(\phi) = N \; M^4 \; w(\chi) \; ,
\ee  
\noindent where $ \phi $ is the inflaton field,
$\chi$ is a dimensionless, slowly varying field 
\be\label{chiflai} 
\chi \equiv \frac{\phi}{\sqrt{N} \;  M_{Pl}}  \; ,
\ee 
\noindent $ w(\chi) \sim \mathcal{O}(1) $ and 
$ M $ is the energy scale of inflation which is determined by the 
amplitude of the scalar adiabatic fluctuations \cite{WMAP} to be
\be\label{Mwmap2}
M \sim 0.00319 \; M_{Pl} = 0.77 \times 10^{16} \; {\rm GeV} \; .
\ee
Following the spirit of the Ginsburg-Landau theory of phase transitions,
the simplest choice is a quartic trinomial for the inflaton potential
\cite{nos1,nos2,1sN}:
\be \label{wxi}
w(\chi)= w_0 \pm \frac12 \; \chi^2 + \frac{h}3 \; \sqrt{\frac{y}2} \; \chi^3 +
\frac{y}{32} \; \chi^4 \; ,
\ee
where the coefficients $ w_0, \; h $ and $ y $ are dimensionless and of order 
one and the signs $ \pm $ correspond to large field and small field inflation, 
respectively (namely, chaotic inflation and new inflation, respectively). $ h $ describes how
asymmetric is the potential, $ y $ is the dimensionless quartic coupling.

Inserting eq.(\ref{wxi}) in eq.(\ref{Vi}) yields,
\be\label{VI}
V(\phi)= V_0 \pm \frac{m^2}{2} \; \phi^2 +  \frac{ m
\; g }{3} \; \phi^3 + \frac{\lambda}{4}\; \phi^4 \; ,
\ee
where the mass $ m^2 $ and the couplings $ \; g $ and $ \lambda $ are given
by the following see-saw-like relations, 
\be 
m = \frac{M^2}{M_{Pl}} \qquad ,  \qquad g = h \; \sqrt{\frac{y}{2 \; N}} 
\; \left( \frac{M}{M_{Pl}}\right)^2  \qquad ,  \qquad \lambda  = 
\frac{y}{8 \; N} \left( \frac{M}{M_{Pl}}\right)^4 \label{acoi} 
\qquad ,  \qquad  V_0 = N \; M^4 \; w_0 \; . 
\ee 
Notice that $ y \sim {\cal O}(1) \sim h $ guarantee that $ g  \sim 
{\cal O}(10^{-6}) $ and $ \lambda  \sim {\cal O}(10^{-12}) $ {\bf without}
any fine tuning as stressed in ref. \cite{1sN}. That is, the smallness
of the couplings here is not a consequence of fine tunning but 
follow directly from the form of the inflaton potential
eq.(\ref{Vi}) and the amplitude of the scalar fluctuations that fixes $ M $ \cite{1sN}.

\medskip

During inflation the inflaton field $ \chi $ slowly runs from its initial value
till its final value $ \chi_{end} $ at the absolute minima of the potential $ w(\chi) $.
We have $ \chi_{end} = 0 $ for chaotic inflation while $ \chi_{end} $ turns
to be a function of the coupling $ y $ and the asymmetry $ h $ for new inflation.
Inflation ends after a {\it finite} number of efolds provided
$$
w(\chi_{end}) = w'(\chi_{end}) = 0
$$
Enforcing this condition in the inflationary potential eq.(\ref{wxi}) determines 
the constant $ w_0 $. We have $ w_0 = 0 $ for chaotic inflation while $ w_0 $ turns
to be a function of the coupling $ y $ and the asymmetry $ h $ for new inflation.
$ \chi $ can vary in the interval $ (0, \infty) $ for chaotic inflation while 
$ \chi $ is in  the interval $ (0,\chi_{end}) $ for new inflation.

We derive explicit formulae and study in detail 
the spectral index $ n_s $ of the adiabatic fluctuations, the ratio $ r $
of tensor to scalar fluctuations and the running index $ d n_s/d 
\ln k $ both for trinomial chaotic inflation and trinomial new inflation. 

\medskip

The small coupling limit $ y \to 0 $ of eqs.(\ref{wxi})-(\ref{VI}) corresponds 
to the quadratic monomial potential while the strong coupling limit $ y \to \infty $
yields in chaotic inflation the massless quartic monomial potential. The extreme asymmetric limit
$ |h| \to \infty $ yields a massive model without quadratic term. In such a limit
the product $ |h| \; M^2 $ must be kept fixed since it is determined by the 
amplitude of the scalar fluctuations.

\medskip

In this paper we perform Monte Carlo Markov Chains (MCMC) analysis of the commonly
available CMB and Large Scale Structure (LSS) data. 
For CMB we considered the three years WMAP data, which provide the dominating contribution, 
and also small scale data
(ACBAR, CBI2, BOOMERANG03). For LSS we considered SDSS (Sloan Digital Sky Survey) \cite{SDSS}.  
We used the CosmoMC program \cite{mcmc} within the effective field theory of inflation.
We used a large collection of parallel chains with a total number of
  samples close to five million.
For chaotic and new inflation we imposed as a hard constraint that the spectral index 
$ n_s $ of the adiabatic fluctuations and the ratio $ r $ are given by the
analytic formulas at order $ 1/N $ we derived for the trinomial inflaton potential. 
Our analysis differs in this {\bf crucial} aspect from previous MCMC studies involving the WMAP3
data set [11]. As natural within inflation, we also included the inflationary
consistency relation $ n_T = -r/8 $ on the tensor spectral index. 
This constraint is in any case practically negligible. 
The details of the MCMC analysis are explained in Section V. 

\medskip

Our approach is different to the inflationary flow equations \cite{flujo}
where the inflaton potential {\bf changes} (that is,  {\bf the model changes}) as the flow goes on.
We work with a {\bf given} potential within the Ginsburg-Landau (GL) spirit, 
that is the trinomial potential.
We investigate the physics of the chosen potential in the parameter space driven by 
the data through the Monte Carlo Markov Chains.
In our work, $ n_s $ and $ r $ are computed analytically to order $ 1/N $ for
the trinomial potentials [eqs.(\ref{nscao})-(\ref{rcao}) and (\ref{nstrino})-(\ref{rnue})].
Since $ N \sim 50 $, higher order corrections in $ 1/N $ are irrelevant and can be safely neglected. 
As shown in ref. \cite{1sN}, the slow roll expansion is in fact a systematic expansion in $ 1/N $. 

\medskip

We allowed seven cosmological parameters to vary in our MCMC runs:
the scalar spectral index $ n_s $, the tensor-scalar ratio $r$,
the baryonic matter fraction $ \omega_b $, the dark matter fraction
$ \omega_c $, the optical depth $ \tau $, the ratio of the (approximate) sound
horizon to the angular diameter distance $ \theta $ and the primordial
superhorizon power in the curvature perturbation at $0.05~{\rm Mpc}^{-1}$,
$ A_s $. We allowed the same seven parameters to vary in the MCMC runs for chaotic
and new inflation. 

In the case of new inflation, since the characteristic banana--shaped allowed
region in the  $(n_s,r)$ plane [fig. \ref{ene}] is quite narrow and non--trivial, 
it is convenient to use the two independent
variables $ z \equiv \frac{y}8 \; \chi^2 $ and $ h $ in trinomial inflationary setup as MC parameters.
That is, we used the analytic expressions we found at order $ 1/N $,  
to express  $ n_s $ and $ r $ in terms of  $ z $ and $ h $.

\medskip

Concerning priors, we kept the same, standard ones, of the $\Lambda$CDM+$r$
model for the first five parameters ($ \omega_b, \; \omega_c, \; \tau, \;
\theta $ and $ A_s $), while we considered all the possibilities for $z$
and $h$. 

In the case of chaotic inflation we kept $ n_s $ and $ r $ as MC parameters,
imposing as hard priors that they lay in the region described by 
chaotic inflation [fig. 2]. This is technically convenient, since this
region covers the major part of the probability support of $ n_s $ and $ r $ in the
$\Lambda$CDM+$r$ and the parametrization eqs.(\ref{nscao})-(\ref{rcao}) in terms of the
parameters $ z $ and $ h $ becomes quite singular in the limit $h \to -1$. This is
indeed  the limit which allows to cover the region of highest likelihood. 
The priors on the other parameters 
where the same of the $\Lambda$CDM+$r$ model and of new inflation. 

\medskip

We did not marginalize over the SZ amplitude, 
and we did not include non-linear effects in the evolution of 
the matter spectrum. The relative corrections are
in any case not significant [5], especially in the present context.

\medskip

In all our MCMC runs we keep fixed the number of efolds $ N $ since horizon
exit till the end of inflation. 
The reason is that the main physics that determines the value of $ N $ is {\bf not} 
contained in the available data but involves the reheating
era. Therefore, it is {\bf not} reliable to fit $ N $ solely to the CMB and LSS data.
The precise value of $ N $ is certainly near $ N =50 $ [14, 15]. We take here the
value $ N =50 $ as a reference baseline value for numerical analysis, 
but from the explicit expressions obtained in the slow roll $ 1/N $ expansion we see that
both $ n_s - 1 $ and $ r $ scale as  $ 1/N $. Therefore, varying $ N $
produces a scale trasformation in the $(n_r-1,r)$ plane, thus displacing
the black and red curves in fig.  \ref{ene} towards up and left or towards down and
right.  This produces however small quantitative changes in our bounds for $
r $ as well as in the most probable values for $ r $ and $ n_s $. 
MCMC simulations with variable $N$ and imposing
the trinomial new inflation potential yielded $ N \sim 50 $ as the most probable value.

\medskip

We plot in fig. \ref{ene} $ \; r $ vs. $ n_s $ both for chaotic and new
inflation for fixed values of the asymmetry $ h $ and the coupling $ y $
varying along the curves.  We see that the regions of the trinomial new
inflation and chaotic inflation are {\bf complementary} in the $(n_s,r)$
plane.  The red curves are those of chaotic inflation, while black curves
are for new inflation.  The leftmost black line corresponds to new
inflation with a symmetric potential $ h = 0 $. The rightmost black line
describes the case of new inflation with an extreme asymmetric potential $
|h| \gg 1 $. This last line is the border of the region on the right
described by chaotic inflation. Although chaotic inflation covers a much
wider area than new inflation, this wide area is only a small corner of the
parameter space (field $ z \equiv \frac{y}8 \; \chi^2 $, asymmetry $ h $)
as shown by fig. \ref{pdfzh}.

\bigskip

For the trinomial {\bf new inflation} model we find a {\bf lower bound} on $ r $:
\be\label{cotrI}
r > 0.016 \quad (95\% \; {\rm CL}) \quad ,\quad r > 0.049 \quad (68\% \; {\rm CL}) 
\quad {\rm (new \; inflation}) \; .
\ee
while for $ n_s $ we find:
$$
n_s > 0.945 \quad (95\%  \;  {\rm CL}) \quad {\rm (new \; inflation}) \; .
$$
The most probable values are (see fig. \ref{z1yhnsr}),
\be \label{valopt}
n_s \simeq 0.956 \quad ,\quad r\simeq 0.055 \quad {\rm (new \; inflation}) \; .
\ee
For trinomial new inflation there exists the theoretical
upper limits: $ n_s \leq 0.9615\ldots , \; r \leq 0.16$ \cite{nos2}.  Thus,
the {\em most probable value} of $ n_s $ for trinomial new inflation eq.(\ref{valopt}) is
very close to its {\em theoretical limiting value}, and that of $ r $ is not
too far from it (see also fig. \ref{z1yhnsr}).

\medskip

The probability distribution for the asymmetry parameter 
$ h $ is peaked at $ h = 0 $ with
$$
|h| < 4.92 \;  {\rm with} \; 95 \% \; {\rm CL} \quad {\rm new \; inflation} \; .
$$
That is, we find that the most probable trinomial new inflation potential
is symmetric and has a moderate nonlinearity with the quartic coupling $ y
\simeq 2 $ for $ h \simeq 0 $. This is the following potential: 
\be
\label{wopt} w(\chi) = \frac{y}{32} \left(\chi^2 - \frac8{y}\right)^{\! 2}
\; .  
\ee 
The $ \chi \to - \chi $ symmetry is here broken since the
absolute minimum of the potential is at $ \chi \neq 0 $.

\bigskip

For  trinomial {\bf chaotic inflation}, the chaotic symmetric trinomial potential $ h = 0 $ is
almost certainly {\bf ruled out} since $ h < -0.7 $ at 95 \% confidence level (see fig. \ref{pdfzh}).

We see the maximun probability for strong asymmetry $ h < -0.95 $ 
and significant nonlinearity $ 4.207\ldots < y < +\infty $
That is, in chaotic inflation {\em all three} terms in the trinomial potential 
$ w(\chi) $ {\bf do} contribute.

\bigskip

We have not introduced the running of the spectral index $ dn_s / d\ln k $
in our MCMC fits since the running must be very small of the order $ {\cal
  O}(N^{-2}) \sim 0.001 $ in slow-roll and for generic potentials
\cite{1sN}.  We find that adding the new parameter $ dn_s / d\ln k $ to the
MCMC analysis make negligible changes on the fit of $ n_s $ and $ r $. All
this suggests that the present data are {\bf not} yet precise enough to
allow a determination of $ dn_s / d\ln k $.  

\bigskip

In summary, the data favour the breaking of the $ \chi \to -  \chi $ symmetry
both for new and for chaotic inflation. 

Trinomial chaotic inflation produces its best fit to the data in a {\bf very narrow}
corner of the parameter space. 

The potential in chaotic inflation has a single minimum. But this minimum for the
best fit potential happens to be in the boundary
of the parameter space, precisely where the asymmetry of the potential is so large that arises 
an extra minimum to the potential. 
That is, the data are begging the potential to be a double well (that is, a new inflation potential). 
The MCMC runs go towards a potential in the {\em boundary} of the parameter space
and maximal symmetry breaking. This limiting potential exhibits an inflexion point
and a significative nonlinearity. 
This strongly suggest that the true potential may be outside the parameter space of chaotic inflation. 
On the contrary, starting with a double well, that is new inflation,
the data are perfectly well fitted with small or even zero asymmetry.

In chaotic inflation, the MCMC minimization leads to a large nonlinearity (from the higher 
order cubic and cuartic terms), making chaotic inflation not stable under higher
order terms in the potential. Chaotic inflation thus results contrary to the Ginsburg-Landau 
spirit in which higher order terms of the potential must be smaller and smaller giving 
irrelevant corrections.

On the contrary, new inflation results perfectly well fitted by the data with very small 
or zero cubic non-linearity, and thus new inflation is stable under higher order terms of 
the potential, which is perfectly well within the Ginsburg-Landau sprit.

\medskip

Indeed, one can fit the WMAP+LSS data both with trinomial new inflation as well as
with trinomial chaotic inflation. However, the trinomial potential that gives the best
fit for chaotic inflation is theoretically {\bf undesirable} for the reasons explained
above. Therefore, trinomial new inflation is the {\bf best model} that describes the data.
This implies that we {\bf do have} a lower bound on $ r $ as given by eq.(\ref{cotrI}). 
Moreover, the best fit is obtained with the double well inflation potential eq.(\ref{wopt}) 
yielding the values in eq.(\ref{valopt}) for $ n_s $ and $ r $. 

\bigskip

This paper is organized as follows: in section II we discuss the slow-roll approach as the $ 1/N $
expansion. In sections III and IV we present the trinomial chaotic inflation and the new trinomial 
inflation, respectively, displaying the analytic formulas for the spectral index $ n_s $ of the 
adiabatic fluctuations, the ratio $ r $ of tensor to scalar fluctuations and the running  
index $ d n_s/d \ln k $. In section V we explain our MCMC analysis, we present our MCMC results for 
new and chaotic trinomial inflation and we state our conclusions.

\section{The Inflaton Potential and the $ 1/N $ Slow Roll Expansion}

The description of cosmological inflation is based on an isotropic
and homogeneous geometry, which assuming flat spatial sections is
determined by the invariant distance
\be
ds^2= dt^2-a^2(t) \; d\vec{x}^{\, 2} \label{FRW} \; .
\ee
The scale factor obeys the Friedman equation
\be \label{ef}
H^2(t) = \frac{\rho(t)}{3 \; M^2_{Pl}}  \quad , \quad {\rm where}  \quad 
H(t) \equiv \frac{1}{a(t)} \; \frac{da}{d t} \;  ,
\ee
where $ \rho(t) $ is the total energy density and 
$ M_{Pl}= 1/\sqrt{8\pi G} = 2.4\times 10^{18}\,\textrm{GeV} $.

In single field inflation the energy density is dominated by a
homogeneous scalar \emph{condensate}, the inflaton, whose dynamics
is described by an  \emph{effective} Lagrangian
\be\label{lagra}
{\cal L} = a^3({ t}) \left[ \frac{{\dot
\phi}^2}{2} - \frac{({\nabla \phi})^2}{2 \;  a^2({ t})} - V(\phi) \right]
\; .
\ee
\noindent The inflaton potential $ V(\phi) $ is a slowly varying
function of $ \phi $ in order to permit a slow-roll solution for
the inflaton field $ \phi(t) $.
We showed in ref.\cite{1sN} that combining the slow roll expansion
with the WMAP data yields an inflaton potential of the form 
\be \label{V} 
V(\phi) = N \; M^4 \; w(\chi)  \; ,
\ee  
\noindent where $\chi$ is a dimensionless, slowly varying field 
\be\label{chifla} 
\chi = \frac{\phi}{\sqrt{N} \;  M_{Pl}}  \; ,
\ee 
$w(\chi) \sim \mathcal{O}(1)~,~N \sim 50$ is the number of efolds
since the cosmologically relevant modes exited the horizon till the 
end of inflation and $ M $ is the energy scale of inflation

\medskip

The dynamics of the rescaled field $ \chi $ exhibits the slow
time evolution in terms of the \emph{stretched}
dimensionless time variable, 
\be \label{tau} 
\tau =  \frac{t \; M^2}{M_{Pl} \; \sqrt{N}}  \quad , \quad  
{\cal H} \equiv \frac{H}{\sqrt{N} \; m} = {\cal O}(1) \; .
\ee 
The rescaled variables $ \chi $ and $ \tau $ change slowly with time. 
A large change in the field amplitude $\phi$ results in a small change 
in the $ \chi $ amplitude, a change in $\phi \sim  M_{Pl}$ results in a 
$ \chi $ change $ \sim 1/\sqrt{N} $. The form of the potential, eq.(\ref{V}) 
and the rescaled dimensionless inflaton field eq.(\ref{chifla}) and time 
variable $ \tau $ make {\bf manifest} the slow-roll expansion as a consistent 
systematic expansion in powers of $ 1/N $ \cite{1sN}.  

\medskip

The inflaton mass around the minimum is given by a see-saw formula
$$
m = \frac{M^2}{M_{Pl}}  \sim 2.45 \times 10^{13} \, \textrm{GeV} \; .
$$
The Hubble parameter when the cosmologically relevant modes exit the horizon
is given by
$$
H  = \sqrt{N} \; m \, {\cal H} \sim 1.0 \times 10^{14}\,\textrm{GeV}
= 4.1 \; m \; ,
$$
where we used that $  {\cal H} \sim 1 $. As a result, $ m\ll M $ and 
$ H \ll M_{Pl} $. 

\medskip

A Ginsburg-Landau realization of the inflationary potential 
that fits the amplitude of the CMB anisotropy 
remarkably well, reveals that the
Hubble parameter, the inflaton mass and non-linear couplings are
see-saw-like, namely  powers of the ratio $M/M_{Pl}$ multiplied by
further powers of $ 1/N $. Therefore, the smallness of the couplings is not 
a result of fine tuning but a {\bf natural} consequence of the form of
the potential and the validity of the effective field theory
description and slow roll. The quantum expansion in loops is
therefore a double expansion on $ \left(H/M_{Pl}\right)^2 $ and $
1/N $. Notice that graviton corrections are  also at least of
order $ \left(H/M_{Pl}\right)^2 $ because the amplitude of tensor
modes is of order $H/M_{Pl}$ . We showed that the form of the potential which 
fits the WMAP data and is consistent with slow roll 
eqs.(\ref{V})-(\ref{chifla}) implies the small values for the inflaton 
self-couplings \cite{1sN}. 

\medskip

The equations of motion in terms of the dimensionless  rescaled field 
$ \chi $ and the slow time variable $ \tau $ take the form,
\bea \label{evol} 
&&  {\cal H}^2(\tau) = \frac13\left[\frac1{2\;N} 
\left(\frac{d\chi}{d \tau}\right)^2 + w(\chi) \right] \quad , \cr \cr
&& \frac1{N} \;  \frac{d^2
\chi}{d \tau^2} + 3 \;  {\cal H} \; \frac{d\chi}{d \tau} + w'(\chi) = 0 \quad .
\eea 
The slow-roll approximation follows by neglecting the
$ \frac1{N} $ terms in eqs.(\ref{evol}). Both
$ w(\chi) $ and $ {\cal H}(\tau) $ are of order $ N^0 $ for large $ N $. Both
equations make manifest the slow roll expansion as an expansion in $ 1/N $.

The number of e-folds $ N[\chi] $ since the field $ \chi $ exits the horizon 
till the end of inflation (where $ \chi $ takes the value $ \chi_{end} $) 
can be computed in close form from eqs. (\ref{evol}) in the slow-roll 
approximation (neglecting $ 1/N $ corrections)
\be \label{Nchi}
\frac{N[\chi]}{N} = -\int_{\chi}^{\chi_{end}}  \;
\frac{w(\chi)}{w'(\chi)} \; d\chi \;  \leqslant 1 \; .
\ee

\medskip

Inflation ends after a finite number of efolds provided
\be\label{condw}
w(\chi_{end}) = w'(\chi_{end}) = 0
\ee
This condition will be enforced in the inflationary potentials.

\medskip

The amplitude of adiabatic scalar perturbations is expressed as
\cite{libros,WMAP,1sN,hu}
\be \label{ampliI}
|{\Delta}_{k\;ad}^{(S)}|^2  = \frac{N^2}{12 \, \pi^2} \;
\left(\frac{M}{M_{Pl}}\right)^4 \; \frac{w^3(\chi)}{{w'}^2(\chi)} \; .
\ee
The spectral index $ n_s $,  its running and the ratio of tensor to scalar 
fluctuations are expressed as
\bea \label{indi}
&&n_s - 1 = -\frac3{N} \; \left[\frac{w'(\chi)}{w(\chi)} \right]^2
+  \frac2{N}  \; \frac{w''(\chi)}{w(\chi)} \quad , \cr \cr 
&&\frac{d n_s}{d \ln k}= - \frac2{N^2} \; \frac{w'(\chi) \;
w'''(\chi)}{w^2(\chi)} - \frac6{N^2} \; \frac{[w'(\chi)]^4}{w^4(\chi)}
+ \frac8{N^2} \; \frac{[w'(\chi)]^2 \; w''(\chi)}{w^3(\chi)}\quad , \cr \cr 
&&r = \frac8{N} \; \left[\frac{w'(\chi)}{w(\chi)} \right]^2 \quad .
\eea
In eqs.(\ref{Nchi})-(\ref{indi}) the field $ \chi $ is computed at horizon 
exiting. We choose $ N[\chi] = N = 50 $.

Since, $ w(\chi) $ and  $ w'(\chi) $ are of order one, we
find from eq.(\ref{ampliI})
\be\label{Mwmap}
\left(\frac{M}{M_{Pl}}\right)^2 \sim \frac{2
\, \sqrt{3} \, \pi}{N} \; |{\Delta}_{k\;ad}^{(S)}| \simeq  1.02
\times 10^{-5} \; .
\ee
where we used $ N \simeq 50 $ and the WMAP
value for $ |{\Delta}_{k\;ad}^{(S)}| = (4.67 \pm 0.27)\times
10^{-5} $ \cite{WMAP}. This fixes the scale of inflation to be
$$
M \simeq 3.19 \times 10^{-3} \; M_{PL} \simeq 0.77
\times 10^{16}\,\textrm{GeV} \; .
$$
This value {\em pinpoints the scale of the potential} during inflation to be at
the GUT scale suggesting a deep connection between inflation and the
physics at the GUT scale in cosmological space-time.

\medskip

We see that $|n_s -1|$ as well as  the ratio $ r $ turn out to be of order 
$ 1/N $. This nearly scale invariance is a natural property of
inflation which is described by a quasi-de Sitter space-time geometry.
This can be understood intuitively as follows: 
the geometry of the universe is scale invariant during de Sitter stage 
since the metric takes in conformal time the form 
$$
ds^2 = \frac1{(H \; \eta)^2}\left[ (d \eta)^2 - (d \vec x)^2 \right] \; .
$$
Therefore, the primordial power generated is scale invariant except
for the fact that inflation is not eternal and lasts for $N$ efolds.
Hence, the  primordial spectrum is scale invariant up to $ 1/N $ 
corrections. The Harrison-Zeldovich values $ n_s = 1, \; r = 0 $ and $ d n_s/d \ln k
= 0 $ correspond to a critical point as discussed in ref.\cite{1sN}.
This gaussian fixed point is {\bf not} the inflationary model that reproduces the data
but the inflation model hovers around it 
in the renormalization group sense with an almost scale invariant spectrum 
of scalar fluctuations during the slow roll stage.

\medskip

We analyze in the subsequent sections chaotic inflation and new inflation
in its simple physical realizations within the Ginzburg-Landau approach
(the trinomial potential)\cite{1sN,nos1,nos2}. We perform with these models
Monte Carlo Markov chains analysis of the three years WMAP data and LSS data.

\section{Trinomial Chaotic Inflation: 
Spectral index, amplitude ratio, running index and limiting cases}\label{tci}

We consider now the trinomial potential with  unbroken symmetry investigated 
in ref.\cite{nos1}:
\be\label{VC}
V(\phi)=  \frac{m^2}{2} \; \phi^2 + \frac{ m \; g }{3} \; \phi^3 + 
\frac{\lambda}{4}\; \phi^4 \; ,
\ee
where $ m^2 > 0 $ and $ g $ and $ \lambda $ are dimensionless couplings.

The corresponding dimensionless potential $ w(\chi) $ eqs.(\ref{Vi})-(\ref{chiflai} )
takes the form
\be\label{trinoC}
w(\chi) = \frac12 \; \chi^2 + \frac{h}3 \; \sqrt{\frac{y}2} \; \chi^3 +
\frac{y}{32} \; \chi^4 \; ,
\ee
where the quartic coupling $ y $ is dimensionless as well as 
the asymmetry parameter $ h $. The couplings in eq.(\ref{VC}) and 
eq.(\ref{trinoC}) are related by eq.(\ref{acoi}).

Chaotic inflation is obtained by choosing the initial field $ \chi $ in 
the interval $ (0,+\infty) $. The inflaton  $ \chi $ slowly rolls down the slope
of the potential from its initial value till the absolute minimum of the 
potential at the origin.

\begin{figure}
\includegraphics{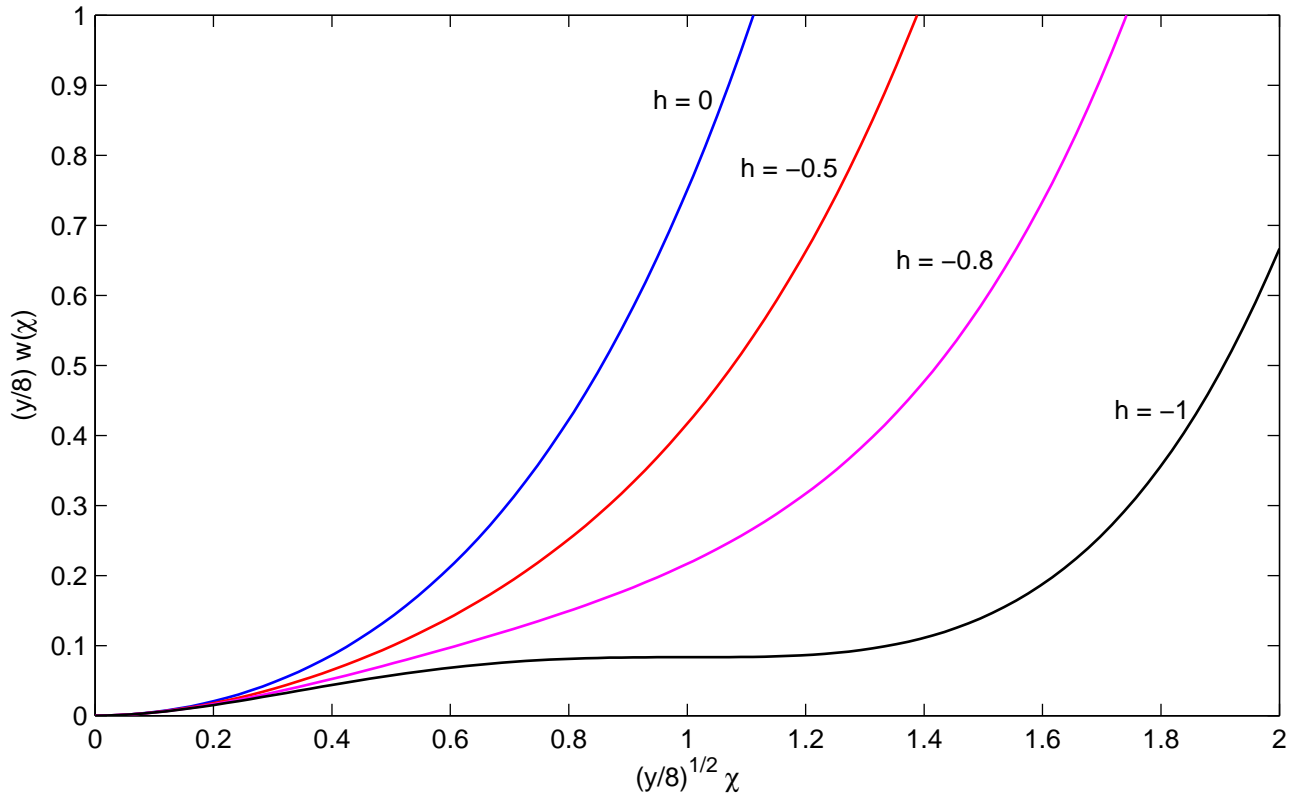}
\caption{Trinomial Chaotic Inflation. We plot here the chaotic inflation trinomial
potential  [eq.(\ref{trinoC}) with positive quadratic term] $ \frac{y}8 \; w(\chi) $ vs.
the field variable $ \sqrt{z}=\sqrt{y/8} \; \chi $ for different values of the asymmetry parameter
$ h $, namely, $ h =0, \; -0.5, \; -0.8 $ and $ -1 $. 
Notice the inflection point at $ \sqrt{z}= 1 $ when $ h=-1 $.}
\label{vv}
\end{figure}

Computing the number of efolds from eq.(\ref{Nchi}), we find the field 
$ \chi $ at horizon crossing related to the couplings $ y $ and $ h $.
Without loss of generality, we choose $ h < 0 $ and shall
work with positive fields $ \chi $. 

The potential eq.(\ref{trinoC}) has extrema at $ \chi = 0 $ and $ \chi_{\pm} $ given by,
\be\label{minD}
\chi_{\pm} = \sqrt{\frac8{y}}\left[ -h \pm i \; \Delta \right] 
\quad , \quad \Delta \equiv \sqrt{1-h^2} \; .
\ee
That is,  for $ |h| < 1 , \; w(\chi) $ has 
only one real extremum at $ \chi = 0 $ while for  $ |h| \geq 1 , \; w(\chi) $ has
three real extrema. There is always a minimum
at  $ \chi = 0 $ since $ w''(0) = 1 $.  At the non-zero extrema we have
$$
w''(\chi_{\pm}) = -2 \; \Delta \; \left(\Delta \pm i \; h \right) \; .
$$
We have for  $ |h|>1 $,
$$
\chi_{\pm} = \sqrt{\frac8{y}}\left[ -h \pm \sqrt{h^2-1} \right] \; , \quad {\rm and}
\quad w''(\chi_{\pm}) =2 \,  \sqrt{h^2-1}\left(\sqrt{h^2-1} \mp h \right) \; .
$$
Hence, for any $ h < -1 $, we have $ w''(\chi_{+}) > 0 $ while $ w''(\chi_{-}) < 0 $.
Notice that $ \chi_{\pm} > 0 $  for $ h < -1 $.

Therefore, we have chaotic inflation for positive field $ \chi $ in the regime $ |h| < 1 $
using the inflaton potential eq.(\ref{trinoC}).

We can also have  chaotic inflation with the potential  eq.(\ref{trinoC})
for negative field if $  h > 3/\sqrt{8} $, but for $  |h| > 3/\sqrt{8} $
the absolute minimum is no more at  $ \chi = 0 $ but at $ \chi_{+} $.
Since $ w(\chi_{+}) < 0 $ for $  |h| > 3/\sqrt{8} $ we have to subtract
in this case the value  $ w(\chi_{+}) $ from  $ w(\chi) $ 
in order to enforce eq.(\ref{condw}).

We consider in subsections \ref{prih}, \ref{potcha} and \ref{HZe} the regime $ -1 < h \leq 0 ,
\; \chi \geq 0 $. The case $ h < -1 $ is analyzed in subsection \ref{hmemu}.

\subsection{The small asymmetry regime: $ -1 < h \leq 0 $.}\label{prih}

In chaotic inflation the inflaton field slowly rolls down the slope
of the potential from its initial value till the absolute minimum of the 
potential at $ \chi = 0 $.

\medskip

It is convenient to define the field variable $z$ by:
\be\label{defz}
z \equiv \frac{y}8 \; \chi^2 \; .
\ee
In terms of $ z $ the chaotic trinomial potential takes the form
$$
w(\chi) = \frac4{y} \; z \; \left(1+\frac43 \, h \; \sqrt{z} + \frac12 \, z \right) \; .
$$
When $ z \lesssim 1 $ we are in the quadratic regime where $ w(\chi) $ is approximated
by the $ \chi^2 $ term. For $ z \gtrsim 1 $ we go to the non-linear regime in $ z $
and all three terms in $ w(\chi) $ are of the same order of magnitude.

\medskip

In fig. \ref{vv}, we plot $ \frac{y}8 \; w(\chi) $ as a function of $ \sqrt{z} $ for several
values of $ h \geq - 1 $. We see that the potential becomes flatter as $ h $ decreases.
For $ h = - 1 $, both $ w'(\chi) $ {\bf and } $ w'(\chi) $ vanish at
$ \sqrt{z} = 1 $. The case $ h = - 1 $ is singular since the inflaton gets stuck
an infinite amount of time at the point $ \sqrt{z} = 1 $.

\medskip

By inserting eq.(\ref{trinoC}) for $ w(\chi) $ into  eq.(\ref{Nchi}) for $ N[\chi] $
and setting $ N[\chi] = N $ we obtain the field $ \chi $ or equivalently $ z $ , 
at horizon exit, in terms of the coupling $ y $ and the asymmetry parameter $ h $,
\be\label{ctrino}
y = z + \frac43 \; h \; \sqrt{z} + \left(1 - \frac43 \; h^2\right) \;
\log\left(1+2 \, h \; \sqrt{z} + z \right) 
 -\frac{4\,h}{3 \, \Delta} \left(\frac52 - 2 \, h^2\right)
\left[\arctan\left(\frac{h+\sqrt{z}}{\Delta}\right)
-\arctan \left(\frac{h}{\Delta}\right)\right] 
\ee
This defines the field $ z $ as a monotonically increasing function of the coupling $ y $
for $ 0 < y, \; z < +\infty $. Recall that $ \chi $ and $ z $ corresponds
to the time of horizon exit.

\medskip

We obtain from eqs.(\ref{ampliI}), (\ref{indi}) and (\ref{trinoC}) the 
spectral index, its running, the ratio $ r $ and the amplitude of adiabatic 
perturbations,
\bea\label{nscao}
n_s&=&1- \frac{y}{2 \, N \; z}\left[ 3 \; \frac{\left(1+2 \, h \; \sqrt{z} + z \right)^2}{
\left(1+\frac43 \, h \; \sqrt{z} + \frac12 \, z \right)^2}-
\frac{1+4 \, h \; \sqrt{z} + 3 \, z}{1+\frac43 \, h \; \sqrt{z} + 
\frac12 \, z}\right] \; , \\ \cr\cr
\frac{d n_s}{d \ln k} &=& \frac{y^2}{2 \,  z^2 \; N^2}\left[ - 
\frac{\left(1+2 \, h \; \sqrt{z} + z \right)\left(h \; \sqrt{z} + \frac32 \, z \right)}
{\left(1+\frac43 \, h \; \sqrt{z} + \frac12 \, z \right)^2} - 3 \; 
\frac{\left(1+2 \, h \; \sqrt{z} + z \right)^4}{
\left(1+\frac43 \, h \; \sqrt{z} + \frac12 \, z \right)^4} + \right. \cr \cr\cr
&+& \left.2 \; \frac{\left(1+2 \, h \; \sqrt{z} + z \right)^2
\left(1+4 \, h \; \sqrt{z} + 3 \, z\right)}{\left(1+\frac43 \, h \; \sqrt{z} + 
\frac12 \, z \right)^3} \right]\; , \label{runcao}\\ \cr\cr
r &=& \frac{4 \, y}{N \; z}\frac{\left(1+2 \, h \; \sqrt{z} + z \right)^2}{
\left(1+\frac43 \, h \; \sqrt{z} + \frac12 \, z \right)^2} \; ,\label{rcao} \\ \cr\cr
\label{ampcao}
|{\Delta}_{k\;ad}^{(S)}|^2  &=& \frac{2 \, N^2}{3 \, \pi^2} \; 
\left(\frac{M}{M_{Pl}}\right)^4 \;
\frac{z^2 \; \left(1+\frac43 \, h \; \sqrt{z} + 
\frac12 \, z \right)^3}{y^2 \; \left(1+2 \, h \; \sqrt{z} + z \right)^2} \; .
\eea
In chaotic inflation, the limit $ z \to 0^+ $ implies
$ y \to 0^+ $ (shallow limit), we have in this limit:
\bea
&&y \buildrel{z \to 0^+}\over= 2  \, z +{\cal O}(z^\frac32) \quad , \quad
n_s\buildrel{y \to 0^+}\over=1- \frac{2}{N}\; ,  \cr \cr
&& r \buildrel{y \to 0^+}\over=\frac{8}{N}\quad , \quad
|{\Delta}_{k\;ad}^{(S)}|^2  \buildrel{y \to 0^+}\over= \frac{N^2}{6 \, \pi^2} \; 
\left(\frac{M}{M_{Pl}}\right)^4 \; .\label{cuad}
\eea
The results in the $ y \to 0^+ $ limit are {\bf independent} of the asymmetry $ h $
and coincide with those for the  purely quadratic monomial potential $ \frac12 \; \chi^2 $.

In the limit $ z \to +\infty $ which implies $ y\to +\infty  $ (steep limit), we have
for fixed $ h > - 1 $,
\bea
&&y \buildrel{z \to +\infty}\over=  z + \frac43 \; h \; \sqrt{z} + 
\left(1 - \frac43 \; h^2\right) \; \log z +{\cal O}(1) \; , \cr \cr
&&n_s \buildrel{y \to +\infty}\over=1- \frac{3}{N}\left[1 + \frac43 \; \frac{h}{\sqrt{z}} + 
\left(1 - \frac43 \; h^2\right) \; \frac{\log z}{z} +{\cal O}\left(\frac1{z}\right)\right] \; ,
\cr \cr
&&r \buildrel{y \to +\infty}\over=\frac{16}{N}\left[1 + \frac43 \; \frac{h}{\sqrt{z}} + 
\left(1 - \frac43 \; h^2\right) \; \frac{\log z}{z} +{\cal O}\left(\frac1{z}\right)\right]\; . \cr\cr
&&|{\Delta}_{k\;ad}^{(S)}|^2  \buildrel{y \to +\infty}\over= \frac{N^2}{12 \, \pi^2} \; 
\left(\frac{M}{M_{Pl}}\right)^4 \frac{z}{\left[1 + \frac43 \; \frac{h}{\sqrt{z}} + 
\left(1 - \frac43 \; h^2\right) \; \frac{\log z}{z} +{\cal O}\left(\frac1{z}\right)\right]^2}
\; .
\eea
For $ h = 0 , \; n_s $ and $ r $ in the limit $ y\to +\infty  $
coincide with those of the purely quartic monomial potential $ \frac12 \; \chi^4 $:
\be \label{cuar}
n_s = 1 - \frac{3}{N} \quad ,  \quad r = \frac{16}{N} \; .
\ee

\subsection{The flat potential limit $ h \to -1^{+} $} \label{potcha}

We consider chaotic inflation in the regime $ -1 < h \leq 0 , \; \chi \geq 0 $.

At $ h = -1 $ the potential eq.(\ref{trinoC}) exhibits an inflexion point at 
$ \chi_0 \equiv \sqrt{\frac8{y}} $. Namely, $ w'(\chi_0) =  w''(\chi_0) = 0 $ while 
$ w(\chi_0) = 2/(3 \; y ) > 0 $. That is, this happens 
at $ z_0 = \frac{y}8 \; \chi_0^2 = 1 $.

Therefore, for $ h > - 1 $ but very close to $ h = -1 $, the field evolution strongly
slows down near the point $ \chi = \chi_0 $. This strong slow down shows up in 
the calculation of observables when the field  $ \chi $ at horizon crossing
is $ \chi > \chi_0 $, namely, for $ z > z_0 = 1 $. For $ \chi < \chi_0 $, 
that is $ z < 1 $, the slow down of the field evolution will only appear if $ z \simeq 1 $.
Therefore, the limit $ h = -1 $ is singular since the inflaton field gets trapped
at the point $ z = 1 $.

\medskip

Let us derive the $ h \to -1^{+} $ limit of $ y , \; n_s $ and $ r $ from
eqs.(\ref{ctrino})-(\ref{ampcao}) in the regimes $ z < 1 $ and $ z > 1 $, respectively.

\medskip

When $ h \to -1^{+} $ we see from eq.(\ref{minD}) that $ \Delta \to 0 $ and the arguments of
the two arctan  in eq.(\ref{ctrino}) diverge. Hence, the arctan tends to
$ +\pi/2 $ or $ -\pi/2 $. Depending on whether $ z < 1 $ or $ z > 1 $, 
the $ \pi/2 $'s terms cancel out or add, respectively. The special case $ z = 1 $ is investigated
in the next subsection \ref{HZe}.

\medskip

In the  case  $ z < 1 $ we get:
\bea\label{yL}
&&y = z - \frac43  \; \sqrt{z} - \frac23 \; \log\left(1-\sqrt{z} \right) 
+\frac23\frac{\sqrt{z}}{1-\sqrt{z}} \quad , \quad h \to -1^{+}\quad, \quad z<1 \; ,
\\ \cr
n_s&=&1- \frac{y}{2 \, N \; z}\left[ 3 \; \frac{\left(1-\sqrt{z}\right)^4}{
\left(1-\frac43  \; \sqrt{z} + \frac12 \, z \right)^2}-
\frac{\left(1-\sqrt{z} \right) \left(1-3 \, \sqrt{z} \right)}{1-\frac43  \; \sqrt{z} + 
\frac12 \, z}\right] \quad , \quad h \to -1^{+}\quad, \quad z<1 \; , \\ \cr\cr
r &=& \frac{4 \, y}{N \; z}\frac{\left(1-\sqrt{z}\right)^4}{
\left(1-\frac43  \; \sqrt{z} + \frac12 \, z \right)^2} \quad , 
\quad h \to -1^{+}\quad, \quad z<1 \; ,
\label{rcaoL} \cr\cr
|{\Delta}_{k\;ad}^{(S)}|^2  &=& \frac{2 \, N^2}{3 \, \pi^2} \; 
\left(\frac{M}{M_{Pl}}\right)^4 \;
\frac{z^2 \; \left(1-\frac43 \; \sqrt{z} + \frac12 \, z \right)^3}{y^2 \; \left(1-\sqrt{z}\right)^4}
\; .
\eea
In particular, in the regime $ z \to 1^{-} $ we find,
\bea\label{nsrlim}
&&n_s\buildrel{z \to 1^{-}}\over= 1 - \frac4{N} \quad , \quad
r \buildrel{z \to 1^{-}}\over= \frac{96}{N} \; \left(1-\sqrt{z}\right)^3 \to 0 
\quad , \quad h \to -1^{+} \; , \cr\cr
&& y \buildrel{z \to 1^{-}}\over= \frac23\frac{1}{1-\sqrt{z}} \to \infty
\quad , \quad 
|{\Delta}_{k\;ad}^{(S)}|^2 \buildrel{z \to 1^{-}}\over=\left(\frac{N}{8 \, \pi} \;
\frac{M^2}{M_{Pl}^2} \; y \right)^2\label{amplim}
\; .
\eea
That is, in the limit $  h \to -1^{+} , \; z \to 1^{-} $
the ratio $ r $ becomes {\bf very small} while the
spectral index takes the value $ n_s = 0.92 $. The ratio $ r $ tends
to zero in the regime $ z \to 1^{-} , \;  \chi \to \chi_0 = \sqrt{\frac8{y}} $
because  $ w'(\chi_0) = 0 $ and $ r $ is proportional to
$ {w'}^2(\chi) $ according to eq.(\ref{indi}).

The $ z \to 1^{-} $ regime for $  h \to -1^{+} $ is a strong coupling
limit since $ y \to +\infty $ as shown by eq.(\ref{amplim}). 
In addition, eq.(\ref{amplim}) shows that for large
$ y $ one must keep the product $ y \; M^2 $ fixed
since it is determined by the amplitude of the adiabatic perturbations.
We see from eq.(\ref{amplim}) that $ {\tilde M} \equiv \sqrt{y} \; M $ becomes 
the energy scale of inflation in the $ y \to \infty $ limit:
from eq.(\ref{Mwmap2}), $ {\tilde M} \sim 10^{16}$GeV  according to the observed value of 
$ |{\Delta}_{k\;ad}^{(S)}|/N $ displayed in eq.(\ref{Mwmap}), while
$ M $ and $ m $ vanish in the  $ y \to \infty $ limit
$$
M = \frac{\tilde M}{\sqrt{y}} \buildrel{ y \to \infty }\over= 0 \quad ,  \quad
m = \frac{M^2}{M_{Pl}} = \frac{ {\tilde M}^2}{y \; M_{Pl}} \buildrel{ y \to \infty 
}\over= 0 \quad .
$$

\begin{figure}
\includegraphics{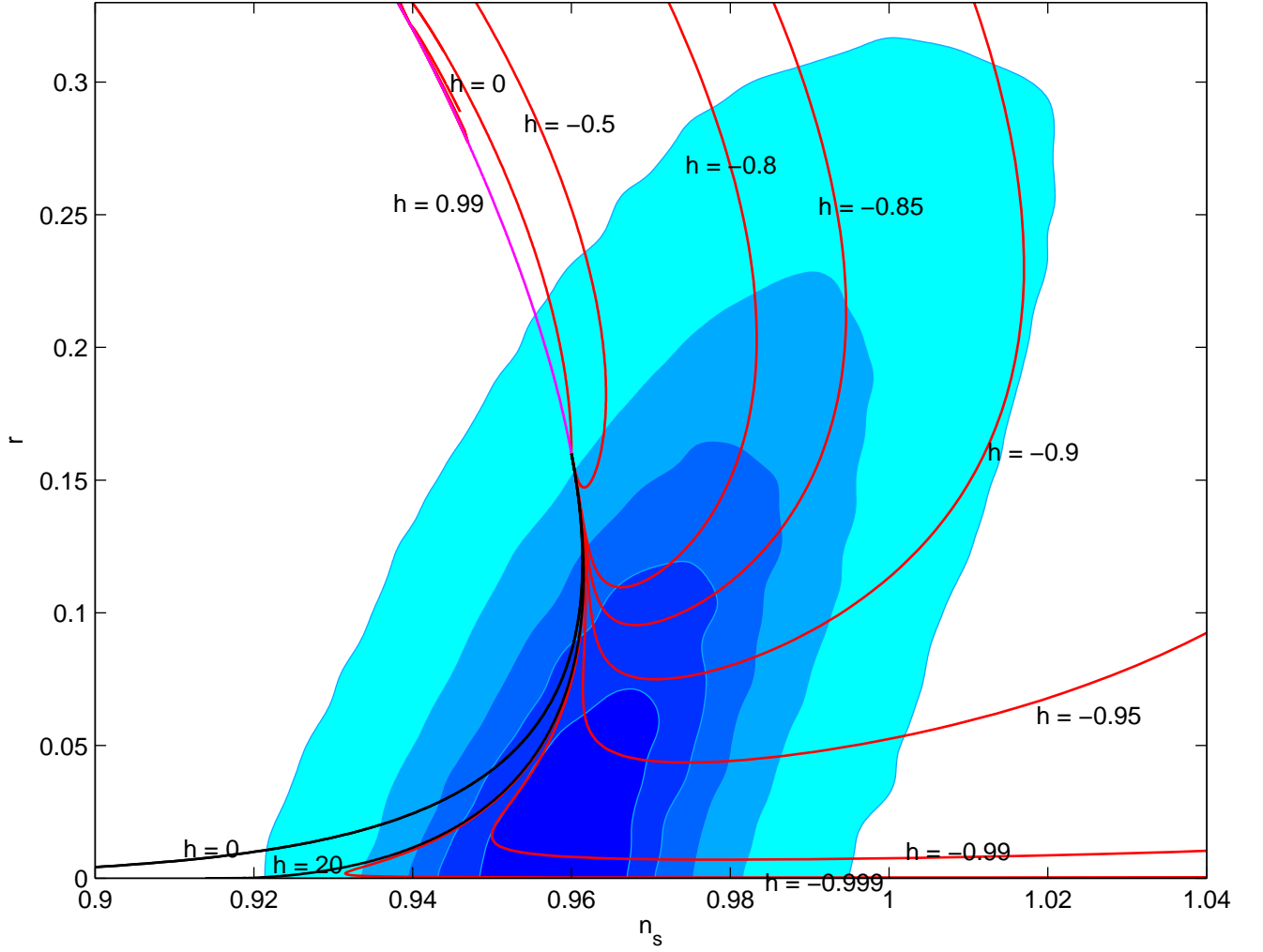}
\caption{Trinomial Inflation. We plot $ r $ vs. $ n_s $ for fixed values of
  the asymmetry parameter $ h $ and the field $ z $ varying
  along the curves. The red curves are those of chaotic inflation with
  $ h\le 0 $ (only the short magenta curve has positive $ h $), while the black
  curves are for new inflation. The color--filled areas correspond to $ 12\%, \; 
  27\%, \;  45\%, \;  68 \% $ and $ 95 \% $ confidence levels according to the WMAP3 and
  Sloan data. The color of the areas goes from the darker to the lighter for increasing CL.
New inflation only covers a narrow area between the black lines while
  chaotic inflation covers a much wider area but, as shown by fig. \ref{pdfzh},
this wide area is only a small corner of the field $ z $ - asymmetry $ h $ plane.
Since new inflation covers the banana-shaped region between the black curves, we see from
this figure that the most probable values of $ r $ are {\bf definitely non-zero} within
trinomial new inflation. Precise
lower bounds for $ r $ are derived from MCMC in eq.(\ref{cotinfr}).}
\label{ene}
\end{figure}

In fig. \ref{ene} we display $ r $ vs. $ n_s $ for fixed values
of the asymmetry parameter $ h $ and the coupling $ y $ varying along the curves.
The red curves correspond to chaotic inflation. 
In the bottom of fig. \ref{ene} we can see the curve for $ h = -0.999 $.
The limiting curve for $ h = -1 $ (not drawn) will reach the point  $ n_s = 0.92 $
and the bottom line $ r = 0 $ as described by eq.(\ref{nsrlim}).

We see from fig. \ref{ene} that the regions of the trinomial new inflation and chaotic inflation are 
{\bf complementary} in the $(n_s,r)$ plane. New inflation describes the region of the $(n_s,r)$ plane
between the two black lines while chaotic inflation describes the whole plane to the right
of the rightmost black line.

\bigskip

In the case $ z > 1 $, we get from eqs.(\ref{ctrino})-(\ref{ampcao}),
\bea\label{yL2}
y &\buildrel{h \to -1^{+}}\over=& \frac{\pi}3 \; \sqrt{\frac2{h+1}} + 
{\cal O}\left([h+1]^0 \right) \to +\infty  \quad, \quad z > 1 \; ,   \cr\cr
n_s &\buildrel{h \to -1^{+}}\over=& 1- \frac{\pi}{3 \, \sqrt{2} \, N \; z}\frac1{\sqrt{h+1}}
\left[ 3 \; \frac{\left(1-\sqrt{z}\right)^4}{
\left(1-\frac43  \; \sqrt{z} + \frac12 \, z \right)^2}-
\frac{\left(1-\sqrt{z} \right) \left(1-3 \, \sqrt{z} \right)}{1-\frac43  \; \sqrt{z} + 
\frac12 \, z}\right] + {\cal O}\left([h+1]^0 \right) \quad, \quad z > 1 \; , \cr \cr\cr
r  &\buildrel{h \to -1^{+}}\over=& \frac{8 \, \pi}{3 \, \sqrt{2}\; N \; z}\frac1{\sqrt{h+1}}
\frac{\left(1-\sqrt{z}\right)^4}{\left(1-\frac43  \; \sqrt{z} + \frac12 \, z \right)^2}  
+ {\cal O}\left([h+1]^0 \right)\quad, \quad z > 1 \; .
\label{rcaoL2} 
\eea
In this strong coupling regime 
the indices become {\bf very large} and hence in contradiction with the data.
In addition, the slow roll expansion cannot be trusted when the coefficients
of $ 1/N $ become large compared with unit. In conclusion, for $ h \to -1^{+} $ 
the case of large field $ z > 1 $ is excluded by the data.

\subsection{The singular limit $ z = 1 $ and then $ h \to -1^{+} $ 
yields the Harrison-Zeldovich spectrum}\label{HZe}

We study in this section the case $ z = 1 $ in trinomial chaotic inflation.
We obtain from eq.(\ref{ctrino})
\be\label{yz1}
y \buildrel{z=1}\over= 1 + \frac43 \; h  + \left(1 - \frac43 \; h^2\right) \;
\log\left[2 \, (1+h)\right] -\frac{4\,h}{3 \, \Delta} \left(\frac52 - 2 \, h^2\right)
\left[\arctan\left(\frac{h+1}{\Delta}\right)
-\arctan \left(\frac{h}{\Delta}\right)\right]  \; .
\ee
For  $ z = 1 $, we find from eqs.(\ref{nscao})-(\ref{ampcao}) and (\ref{yz1}) in the limit
$ h \to -1^{+} $,
\bea
&& y \buildrel{z=1, \; h \to -1^{+}}\over= \frac{\pi}3 \; \sqrt{\frac1{2(h+1)}}
- \frac13 \; \log(h+1) +  {\cal O}\left([h+1]^0\right) \; , \cr \cr
&& n_s \buildrel{z=1, \; h \to -1^{+}}\over= 1 +
\frac{2 \; \pi}{N} \; \sqrt{2 \, (h+1)} \left[1+ {\cal O}\left(h+1\right)\right] \; ,\cr \cr
&& r \buildrel{z=1, \; h \to -1^{+}}\over=\frac{16 \; \pi}{N} \; \sqrt2 \; (h+1)^{\frac32}
\left[1+ {\cal O}\left(h+1\right)\right] \; ,\cr \cr
&&|{\Delta}_{k\;ad}^{(S)}|^2 \buildrel{z=1, \; h \to -1^{+}}\over=
\frac{N^2}{36} \; \left[\frac{M}{\pi \; M_{Pl} \;  (h+1)^{\frac14}}\right]^4 \; .
\eea
Therefore, we reach the Harrison-Zeldovich spectrum $ n_s = 1, \; r = 0 $ as a 
limiting value. This is a strong coupling regime $ y \to \infty $ where in addition
we must keep the ratio $ M/(h+1)^{\frac14} $ fixed for $ h \to -1^{+} $
since  it is determined by the amplitude of the adiabatic perturbations.
That is, we must keep
$$
{\bar M} \equiv  \frac{M}{(h+1)^{\frac14}} \quad {\rm fixed~~ as~~} h \to -1^{+} \; ,
$$
while $ M $ as well as $ m $ go to zero. Actually, the {\bf whole} potential $ V(\phi) $
eq.(\ref{VC}) {\bf vanishes} in this limit since,
\bea
&&m = \frac{M^2}{M_{Pl}} \quad \buildrel{z=1, \; h \to -1^{+}}\over= \quad 
\frac{{\bar M}^2}{M_{Pl}} \; \sqrt{h+1} \to 0 \; , \cr \cr
&& g = h \; \sqrt{\frac{y}{2 \; N}} \; \left(\frac{M}{M_{Pl}}\right)^2  \quad
\buildrel{z=1, \; h \to -1^{+}}\over= \quad - \sqrt{\frac{\pi}{6 \; N}} \; 
\left(\frac{\bar M}{M_{Pl}}\right)^2 \; \left( \frac{1+h}2 \right)^{\frac14} \to 0 \; ,
\cr \cr
&& \lambda  = \frac{y}{8 \; N} \left( \frac{M}{M_{Pl}}\right)^4 
\quad \buildrel{z=1, \; h \to -1^{+}}\over= \quad 
\sqrt{\frac{\pi}{24 \; N}} \; \left(\frac{\bar M}{M_{Pl}}\right)^4 \; \sqrt{\frac{1+h}2}\to 0
\eea
The inflaton field is therefore a {\bf massless} free field at $ z=1 $ and $ h \to -1^{+} $.
This explains why the corresponding spectrum is the scale invariant Harrison-Zeldovich one.
This is clearly a singular limit since one cannot obtain any inflation from an
identically zero potential. Namely, taking the flat limit $ h \to -1^{+} $ in the
spectrum computed for $ z=1 , \; h > -1 $ with a fixed  number of efolds $ N $, yields
the scale invariant Harrison-Zeldovich spectrum.

Notice that here we keep the number of efolds $ N $ {\em fixed} which makes
the potential to vanish since otherwise the field will be stuck at the 
point  $ z=1 $ when $ h = -1 $ leading to eternal inflation.

In summary, this shows theoretically that the Harrison-Zeldovich spectrum is highly unplausible
and unrealistic since it appears only in the singular limit $ z=1 , \; h \to -1^{+} $
where the inflaton potential identically vanishes. 

Recall that one can also get a Harrison-Zeldovich spectrum letting formally
the number of efolds $N$ to infinity in eqs.(\ref{indi}).

\subsection{The high asymmetry $ h < - 1 $ regime.}\label{hmemu}

In order to fulfill the finite number of efolds condition eq.(\ref{condw}) for $ h < -1 $
we have to add a constant piece to the chaotic inflationary potential eq.(\ref{trinoC}).

We therefore consider as inflaton potential in the  $ h < -1 $ regime,
\be\label{trino2}
w(\chi) = \frac12 \; \chi^2 + \frac{h}3 \; \sqrt{\frac{y}2} \; \chi^3 +
\frac{y}{32} \; \chi^4 + \frac2{y} \; G(h)  \; ,
\ee
The absolute minimum of this potential is at
\be
\chi_{+} = \sqrt{\frac8{y}}\left[ -h + D \right] \quad , \quad  D \equiv \sqrt{h^2-1}
\; , \quad h < - 1  \; .
\ee
and we have
\bea\label{gdeh}
&& G(h) \equiv \frac83 \, h^4 - 4 \, h^2 + 1 + \frac83 \, |h| \, D^3  \; , \cr \cr 
&& w''(\chi_{+}) =2 \,  \sqrt{h^2-1}\left(\sqrt{h^2-1} + |h| \right) > 0 \; .
\eea
That is, the inflaton mass squared in units of $ m^2 $ takes the value
$$ 
2 \; D \; ( D + |h| ) \; .
$$
Here, the inflaton field rolls down  the slope
of the potential from its initial value larger than $ \chi_{+} $ till the absolute minimum of the 
potential at $ \chi = \chi_{+} $.

By inserting eq.(\ref{trino2}) for $ w(\chi) $ into  eq.(\ref{Nchi}) for $ N[\chi] $
and setting $ N[\chi] \equiv  N $ we obtain the field $ \chi $ or equivalently $ z  = \frac{y}8 \; \chi^2 $ , 
at horizon exit, in terms of the coupling $ y $ and the asymmetry parameter $ h $:
\bea\label{2trino}
&&y = z + \frac43 \; h \; \sqrt{z} + 1 +\frac23 \; h \; (D-h) 
+ 2 \, G(h) \; \log\frac{\sqrt{z}}{D-h} +
\frac{16}3 \;  h \; (h^2-1) \; (D-h) \; \log\left(\frac{\sqrt{z}+h+D}{2 \, D}\right) \; .
\eea
This defines the field $ z $ as a monotonically increasing function of the coupling $ y $
for 
$$ 
0 < y  < +\infty \quad , \quad  z_+ = (D - h)^2 < z < +\infty \; .
$$
Recall that $ \chi $ and $ z $ corresponds to the time of horizon exit. 

We obtain from eqs.(\ref{ampliI}), (\ref{indi}) and (\ref{trino2}) the 
spectral index,  the ratio $ r $ and the amplitude of adiabatic 
perturbations,
\bea\label{nscao2}
&& n_s=1 - \frac{y}{N}  \frac1{(\sqrt{z}+ h- D)^2}
\left[ \frac{6 \, z \; (\sqrt{z}+ h + D)^2}{\left[z +2 \, (D + \frac{h}3 ) \; \sqrt{z} - \frac23 \; h \; (D-h)-1\right]^2}
- \frac{1 + 4 \, h \; \sqrt{z} + 3 \, z}{z +2 \, (D + \frac{h}3 ) \; \sqrt{z} - \frac23 \; h \; (D-h)-1}
\right]  \; , \cr \cr
&& r =   \frac{16 \; y}{N} \; \frac{z}{(\sqrt{z}+h - D)^2} \; 
\frac{(\sqrt{z}+ h + D)^2}{\left[z +2 \, (D + \frac{h}3 ) \; \sqrt{z} - \frac23 \; h \; (D-h)-1\right]^2} \; , \\ \cr
&&|{\Delta}_{k\;ad}^{(S)}|^2  = \frac{N^2}{12 \, \pi^2} \; 
\left(\frac{M}{M_{Pl}}\right)^4 \;
\frac{\left[G(h) + 2 \, z+ \frac83  \, h  \,  z^{3/2} + 
z^2\right] \; (\sqrt{z}+h - D)^2 \; \left[z +2 \, (D + \frac{h}3 ) \; \sqrt{z} -
 \frac23 \; h \; (D-h)-1\right]^2}{y^2 \; z \; (\sqrt{z}+ h + D)^2}   \; . \nonumber
\eea

\bigskip

When  $ \sqrt{z} \to \sqrt{z_+} , \; y $ vanishes quadratically as,
$$
y \buildrel{z \to z_+}\over= 2 \; 
\left(\sqrt{z} - \sqrt{z_+}\right)^2 + {\cal O} 
\left(\left[\sqrt{z} - \sqrt{z_+}\right]^3\right) \; .
$$
In this limit the spectral index,  the ratio $ r $ and the amplitude of adiabatic 
perturbations become, 
$$
 n_s \buildrel{y \to 0}\over= 1 - \frac2{N} \quad , \quad r \buildrel{y \to 0}\over= \frac8{N} \quad , \quad 
|{\Delta}_{k\;ad}^{(S)}|^2 \buildrel{y \to 0}\over= \frac{N^2}{6 \, \pi^2} \; 
\left(\frac{M}{M_{Pl}}\right)^4 \; .
$$
These results are {\bf independent} of the asymmetry $ h $
and coincide with those for the  purely quadratic monomial potential $ \frac12 \; \chi^2 $.

We see here that
\be\label{regir}
\frac8{N} < r < \frac{16}{N} \quad , \quad 1 - \frac3{N} < n_s < 1 - \frac2{N} 
\quad {\rm for}\quad  0 < y < \infty \; .
\ee
Namely, the regime $ h < - 1 $ of chaotic inflation covers values of $ r $ {\bf larger} than the 
weak coupling limiting value $ r = \frac8{N} $ eq.(\ref{cuad}) and {\bf smaller} than the $ \to \infty $ 
pure quartic potential value $ r = \frac{16}{N} $ eq.(\ref{cuar}).

\section{Trinomial New Inflation: Spectral index, amplitude ratio,
  running index and limiting cases}

We consider here new inflation described by the trinomial potential with broken symmetry 
investigated in ref.\cite{nos1,nos2}
\be\label{VN}
V(\phi)= V_0 - \frac{m^2}{2} \; \phi^2 + \frac{ m \; g }{3} \; \phi^3 + 
\frac{\lambda}{4}\; \phi^4 \; ,
\ee
where $ m^2 > 0 $ and $ g $ and $ \lambda $ are dimensionless couplings.
The corresponding dimensionless potential $ w(\chi) $ takes the form
\be\label{trino}
w(\chi) = -\frac12 \; \chi^2 + \frac{h}3 \; \sqrt{\frac{y}2} \; \chi^3 +
\frac{y}{32} \; \chi^4 + \frac2{y} \; F(h)  \; ,
\ee
where the quartic coupling $ y $ is dimensionless as well as 
the asymmetry parameter $ h $. The couplings in eq.(\ref{VN}) and 
eq.(\ref{trino}) are related by,
\be 
g = h \; \sqrt{\frac{y}{2 \; N}} \; \left( \frac{M}{M_{Pl}}\right)^2  
\qquad ,  \qquad 
\lambda  = \frac{y}{8 \; N} \; \left( \frac{M}{M_{Pl}}\right)^4 
\label{aco} \; ,
\ee 
and the constant $ w_0 $ [see eq.(\ref{wxi})] is related to $ V_0 $ in eq.(\ref{VN}) by
$$
 w_0 \equiv \frac2{y} \; F(h) = \frac{V_0}{N \; M^4 } \; .
$$
The constant $ F(h) $ ensures that $ w(\chi_+) =  w'(\chi_+) = 0 $ 
at the absolute minimum $ \chi = \chi_+ =\sqrt{\frac8{y}} (\Delta + |h|) $ 
of the potential $ w(\chi) $ according to eq.(\ref{condw}). 
Thus, inflation does not run eternally. $ F(h) $ is given by
$$
F(h) \equiv \frac83 \, h^4 + 4 \, h^2 + 1 + \frac83 \, |h| \, \Delta^3 
\quad , \quad \Delta \equiv \sqrt{h^2 + 1} \; .
$$
The parameter $ h $ reflects how asymmetric is the potential.
Notice that  $ w(\chi) $ is invariant under the changes
$ \chi \to - \chi , \;  h \to - h $. Hence, we can restrict
ourselves to a given sign for $ h $. Without loss of 
generality, we choose $ h < 0 $ and shall
work with positive fields $ \chi $ as in sec. \ref{tci}..

We have near the absolute minimum $ \chi = \chi_+ $,
\be \label{cudr}
w(\chi) \buildrel{\chi \to \chi_+}\over= \Delta ( \Delta + |h| ) \; (\chi -\chi_+)^2 +
{\cal O}\left(\sqrt{y}[\chi - \chi_+]^3 \right)  \; ,
\ee
That is, the inflaton mass squared in units of $ m^2 $ takes the value
$$ 
2 \; \Delta ( \Delta + |h| ) \; .
$$
Notice that the  inflaton mass squared takes the analogous value for chaotic inflation in the
$ h <-1 $ regime changing $ D \Longrightarrow \Delta $
while $ F(h) $ differs from $ G(h) $ given by eq.(\ref{gdeh}) only by the
sign of the $ 4 \; h^2 $ term.

Recall that $ y \sim {\cal O}(1) \sim h $ guarantees that $ g  \sim 
{\cal O}(10^{-6}) $ and $ \lambda  \sim {\cal O}(10^{-12}) $ {\it without}
any fine tuning as stressed in ref. \cite{1sN}. 

\begin{figure}[ht]
\begin{turn}{-90}
\includegraphics[height=15cm,width=10cm]{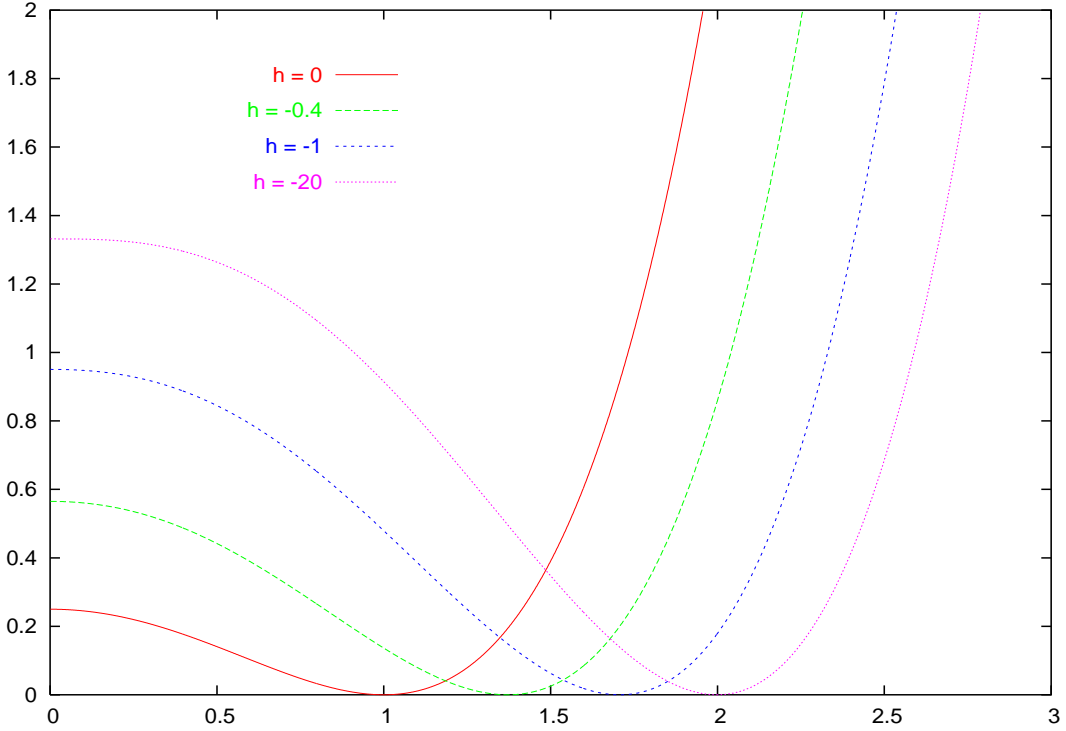}
\end{turn}
\caption{Trinomial New Inflation. The new inflation trinomial potential 
$ \frac{y \; w(\chi)}{8 \; (h^2 + 1)^2} $
[eq.(\ref{trino})] vs. the field variable $ \frac{\sqrt{y} \; \chi}{\sqrt{8} 
\; \sqrt{h^2 + 1}} $ for different
values of the asymmetry parameter $ h = 0, -0.4, -1, -20 $. 
We have normalized here the field variable and the potential
by $h$-dependent factors in order to have a smooth $ |h| \gg 1 $ limit. }
\label{wh}
\end{figure}

In fig. \ref{wh}, we plot 
$$ 
\frac{y}{8 \; (h^2 + 1)^2}  \; w(\chi) 
\quad {\rm as ~ a ~ function ~ of} \quad  
\frac{\sqrt{y}}8  \; \frac{\chi}{\sqrt{h^2 + 1}}
$$ 
for several values of $ h < 0 $. We see that the minimum of the potential 
$$ 
\frac{\sqrt{y} \; \chi_+}{\sqrt{8} \; \sqrt{h^2 + 1}} = 1 + \frac{|h|}{\sqrt{h^2 + 1}} 
$$ 
grows as $ |h| $ grows. Similarly, the maximun of the potential at the origin
$$ 
\frac{y \; w(0)}{8 \; (h^2 + 1)^2} = \frac{F(h)}{4 \; (h^2 + 1)^2} 
$$ 
grows with  $ |h| $.

New inflation is obtained by choosing the initial field $ \chi $ in 
the interval $ (0,\chi_+) $. The inflaton  $ \chi $ slowly rolls down the slope
of the potential from its initial value till the absolute minimum of the 
potential $ \chi_+ $.

Computing the number of efolds from eq.(\ref{Nchi}), we find the field 
$ \chi $ at horizon crossing related to the coupling $ y $ and the asymmetry parameter $ h $.
We obtain by inserting eq.(\ref{trino}) for $ w(\chi) $ into  eq.(\ref{Nchi})
and setting $ N[\chi] = N $,
\bea\label{ntrino}
&& y  = z - 2 \; h^2 -1 - 2  \; |h|  \; \Delta + \frac43 \; 
|h|  \; \left( |h| + \Delta - \sqrt{z} \right) + \cr \cr
&&+\frac{16}{3} \; |h| \;  (\Delta + |h| ) \; \Delta^2  \; 
\log\left[\frac12 \left(1 +  \frac{\sqrt{z} -  |h|}{\Delta}\right)\right] - 
2 \, F(h) \, \log\left[\sqrt{z} \; (\Delta - |h|)\right] 
\quad , \quad z \equiv \frac{y}8 \; \chi^2 \; .
\eea
$ z $ turns to be a monotonically decreasing function of $ y $:
$ z $ decreases from $ z = z _+ = (\Delta + |h|)^2$ till $ z = 0 $ when
$ y $ increases from $ y = 0 $ till $ y = \infty $.
When  $ \sqrt{z} \to \sqrt{z_+} , \; y $ vanishes quadratically as,
$$
y \buildrel{z \to z_+}\over= 2 \; 
\left(\sqrt{z} - \sqrt{z_+}\right)^2 + {\cal O} 
\left(\left[\sqrt{z} - \sqrt{z_+}\right]^3\right) \; .
$$
We obtain in analogous way from eqs.(\ref{ampliI}), (\ref{indi}) and (\ref{trino}) the 
spectral index, its running, the ratio $ r $ and the amplitude of adiabatic 
perturbations,
\bea\label{nstrino}
&& n_s=1 - 6 \,  \frac{y}{N} \, \frac{z \; (z + 2  \, h  \, 
\sqrt{z} -1)^2}{\left[F(h)  -2 \, z+ \frac83  \, h  \,  z^{3/2} +
z^2\right]^2} +  \frac{y}{N} \, \frac{ 3 \, z+ 4 \, h  \, \sqrt{z} -1}{
F(h)- 2 \, z+ \frac83  \, h  \,  z^{3/2} + z^2} \quad , \\ \cr\cr
\label{rtrino}
&&\frac{d n_s}{d \ln k}= - \frac2{N^2} \; \sqrt{z} \; y^2 \; 
\frac{(z + 2  \, h  \, \sqrt{z} -1)(h + \frac32 \; \sqrt{z})}{
\left[F(h)  -2 \, z+ \frac83  \, h  \,  z^{3/2} +z^2 \right]^2} 
- \frac{24}{N^2} \;  y^2 \; z^2 \; \frac{(z + 2  \, h  \, \sqrt{z} -1)^4}{
\left[F(h)  -2 \, z+ \frac83  \, h  \,  z^{3/2} +z^2 \right]^4} \cr\cr
&& + \frac8{N^2} \;  y^2 \; z \; \frac{ (3 \, z+ 4 \, h  \, \sqrt{z} -1)
(z + 2  \, h  \, \sqrt{z} -1)^2}{\left[F(h)  -2 \, z+ \frac83  \, h  \,  
z^{3/2} +z^2 \right]^3} \quad , \label{runnew} \\ \cr \cr
&& r = 16 \,   \frac{y}{N} \, \frac{z \; (z + 2  \, h  \, \sqrt{z} -1
)^2}{\left[F(h)  -2 \, z+ \frac83  \, h  \,  z^{3/2} +z^2
\right]^2}  \quad  , \label{rnue} \\ \cr\cr
&&|{\Delta}_{k\;ad}^{(S)}|^2  = \frac{N^2}{12 \, \pi^2} \; 
\left(\frac{M}{M_{Pl}}\right)^4 \;
\frac{\left[F(h)- 2 \, z+ \frac83  \, h  \,  z^{3/2} + 
z^2\right]^3}{y^2 \; z \; (z + 2  \, h  \, \sqrt{z} -1)^2} \; .\label{dtrino}
\eea

\subsection{The shallow limit (weak coupling) $ y \to 0 $}

From eq.(\ref{ntrino}) we see that in the shallow limit $ y \to 0 , \; z $ tends to 
$ z_+ = (\Delta + |h|)^2 $.
$ y(z) $ has its minimum $ y = 0 $ at $ z =  z_+ $. We find from eqs.(\ref{ntrino})-(\ref{dtrino}), 
\bea\label{cotrino}
&& n_s \buildrel{y \to 0}\over= 1 - \frac{2}{N} \simeq 0.96  
\quad , \quad  r \buildrel{y \to 0}\over= \frac{8}{N}  \simeq 0.16 , \\ \cr
&& \frac{d n_s}{d \ln k} \buildrel{y \to 0}\over= 
-\frac2{N^2}\simeq -0.0008 \quad , \quad 
|{\Delta}_{k\;ad}^{(S)}|^2  \buildrel{y \to 0}\over= \frac{N^2}{3 \, \pi^2} \; 
\left(\frac{M}{M_{Pl}}\right)^4 \; \Delta(\Delta+|h|) \; ,
\eea
which coincide with $ n_s , \; \frac{d n_s}{d \ln k}$ and $ r $ 
for the monomial quadratic potential. That is, the  $ y \to 0 $ limit
is $h$-independent except for $|{\Delta}_{k\;ad}^{(S)}|$.
For fixed $ h $ and $ y \to 0 $ the inflaton potential eq.(\ref{trino}) 
becomes purely quadratic as we see from eq.(\ref{cudr}):
\be \label{maslim}
w(\chi) \buildrel{y \to 0}\over= \Delta ( \Delta + |h| ) \; (\chi -\chi_+)^2 +
{\cal O}(\sqrt{y} )  \; .
\ee
Notice that the amplitude of scalar adiabatic fluctuations eq.(\ref{cotrino})
turns out to be proportional to the square mass of the inflaton in this 
regime. We read this mass squared from eq.(\ref{maslim}): $ 2 \; \Delta ( \Delta + |h| ) $
in units of $ m^2 $. 
The shift of the inflaton field by $ \chi_+ $ has no observable consequences.
For $ h = 0 $ we recover the results of the monomial quadratic potential. 

\subsection{The  steep limit (strong coupling) $ y \to \infty $}

In the  steep limit $ y \to \infty, \;  z $ tends to zero for new inflation.
We find from eq.(\ref{ntrino})
\be\label{trikG}
y  \buildrel{z \to 0}\over=- F(h) \; \log z
-q(h) -1 + {\cal O}(\sqrt{z}) \quad  ,
\ee
where
$$
q(h) \equiv 2 \,  F(h) \log\left(\Delta-|h|\right) -
\frac23 \; \left( h^2 + |h| \; \Delta \right) 
\left\{ 8 \, \Delta^2 \, 
\log\left[\frac12 \left(1 - \frac{|h|}{\Delta}\right)\right] - 1 \right\}
\; ,
$$
$ q(h) $ is a monotonically increasing function of the asymmetry 
$ |h| : \; 0 \leq q(h) < \infty $ for $ 0 < |h| < \infty $.

Then, eqs.(\ref{nstrino})-(\ref{rtrino}) yield,
\bea\label{nsrtrikG}
&&n_s \; \; \buildrel{y \gg 1}\over=1 - \frac{y}{N \; F(h)}
\qquad , \qquad r \; \; \ \buildrel{y \gg 1}\over= 
\frac{16 \; y}{N \; F^2(h)} \, e^{- \frac{y+1+q(h)}{F(h)}}\quad ,
\cr \cr
&& \frac{d n_s}{d \ln k} \; \; \buildrel{y \gg 1}\over= -\frac{2 \; y^2 \; |h|}{N^2 \;
 F^2(h)} \, e^{- \frac{y+1+q(h)}{2 \; F(h)}}\quad , \cr \cr
&& |{\Delta}_{k\;ad}^{(S)}|^2  \; \; \ \buildrel{y \gg 1}\over=  \; \; \
\frac{N^2}{12 \, \pi^2} \; \left(\frac{M}{M_{Pl}}\right)^4 \; 
\frac{F^3(h)}{y^2} \; e^{\frac{y+1+q(h)}{F(h)}} \quad .
\eea
In the case $ h = 0 $  we recover from the steep limit 
eqs.(\ref{trikG})-(\ref{nsrtrikG}) the results for new inflation 
in the large $ y $ regime: we have $ F(0) = 1 $ and $ q(0) = 0 $ and eq.(\ref{nsrtrikG}) becomes,
\bea\label{nsrtrikGh0}
&&n_s \buildrel{y \gg 1, h \to 0}\over=1 - \frac{y}{N}
\quad , \quad r \buildrel{y \gg 1, h \to 0}\over= 
\frac{16 \; y}{N} \, e^{-y-1}\quad ,
\cr \cr
&& \frac{d n_s}{d \ln k}\buildrel{y \gg 1, h \to 0}\over= 
-\frac{2 \; y^2 \; |h|}{N^2} \, e^{-y-1}\quad , \cr \cr
&& |{\Delta}_{k\;ad}^{(S)}|^2  \buildrel{y \gg 1, h \to 0}\over= 
\frac{N^2}{12 \, \pi^2} \; \left(\frac{M}{M_{Pl}}\right)^4 \; 
\frac{e^{y+1}}{y^2} \;  \quad .
\eea
Notice that the slow-roll approximation is no longer valid when the coefficient
of $ 1/N $ becomes much larger than unity.

\subsection{The extremely asymmetric limit $ |h| \to \infty $}

Eqs.(\ref{ntrino})-(\ref{dtrino}) have a finite limit for $ |h| \to \infty 
$ with $ y $ and $ z $ scaling as $ h^2 $. Define,
$$
Z \equiv \frac{z}{h^2} \quad , \quad Y \equiv \frac{y}{h^2} \; .
$$
We have $ 0 \leq Z \leq 4 $ for $ +\infty \geq Y \geq 0 $. 
Then, we find for $ |h| \to \infty $ 
from eqs.(\ref{ntrino})-(\ref{dtrino}) keeping $ Z $ and $ Y $ fixed,
\bea \label{hgra}
&& Y = Z - \frac43 \; \sqrt{Z} - 4 - \frac43 \; \log\frac{Z}4 +
\frac{16}{3 \; \sqrt{Z}}\; , \cr \cr
&& n_s = 1 - 6 \; \frac{Y}{N} \; \frac{Z^2 \; (\sqrt{Z} - 2)^2}{
[\frac{16}3 - \frac83 \; Z^{\frac32} + Z^2 ]^2 } + \frac{Y}{N} \;
\frac{ 3 \; Z - 4 \; \sqrt{Z} }{\frac{16}{3}  - \frac83 \; Z^{\frac32} + Z^2 }
\; , \\ \cr
&&\frac{d n_s}{d \ln k}= - \frac2{N^2} \;  Y^2 \; Z \; 
\frac{(\sqrt{Z} -2)(\frac32 \; \sqrt{Z}-1)}{
\left[\frac{16}3  - \frac83 \,  Z^{3/2} + Z^2 \right]^2} \cr\cr
&&- \frac{24}{N^2} \;  Y^2 \; Z^4 \; \frac{(\sqrt{Z} -2)^4}{
\left[\frac{16}3  - \frac83 \,  Z^{3/2} + Z^2 \right]^4} 
+ \frac8{N^2} \;  Y^2 \; Z^{\frac52} \; 
\frac{(3 \,\sqrt{Z} - 4)(\sqrt{Z} -2)^2}{
\left[\frac{16}3  - \frac83 \,  Z^{3/2} + Z^2 \right]^3}
\quad , \label{runh} \\ \cr \cr
&& r = 16 \; \frac{Y}{N} \; \frac{Z^2 \; (\sqrt{Z} - 2)^2}{
[\frac{16}{3} - \frac83 \; Z^{\frac32} + Z^2 ]^2 }  \; ,  \cr \cr
&& |{\Delta}_{k\;ad}^{(S)}|^2  = \frac{N^2 \; h^2}{12 \, \pi^2} \; 
\left(\frac{M}{M_{Pl}}\right)^4 \;
\frac{[\frac{16}{3}  - \frac83 \; Z^{\frac32} + Z^2]^2}{Y^2 \; Z^2  
\; (\sqrt{Z} - 2)^2} \; . \label{amplihg}
\eea
In the  $ |h| \to \infty $ limit, the inflaton potential takes the form
$$
W(\chi) \equiv \lim_{|h| \to \infty} \frac{w(\chi)}{h^2} =
\frac{32}{3 \; Y} - \frac13 \; \sqrt{\frac{Y}{2}} \; \chi^3 
+ \frac{Y}{32} \; \chi^4 \; .
$$
This is a broken symmetric potential without quadratic term.
Notice that the cubic coupling has dimension of a mass in eq.(\ref{VN})
and hence this is {\bf not} a massless potential contrary to the quartic 
monomial $ \chi^4 $. 

In addition, eq.(\ref{amplihg}) shows that for large 
$ |h| $ one must keep the product $ |h| \; M^2 $ fixed
since this is determined by the amplitude of the adiabatic perturbations.
We see from eq.(\ref{amplihg}) that  in the $ |h| \to \infty $ limit, 
$ {\tilde M} \equiv \sqrt{|h|} \; M $ becomes the energy scale of inflation.
$ {\tilde M} \sim 10^{16}$GeV  [eq.(\ref{Mwmap2})] according to the observed value of 
$ |{\Delta}_{k\;ad}^{(S)}|/N $ displayed in eq.(\ref{Mwmap}), while
$ M $ and $ m $ vanish as $ |h| \to \infty $,
$$
M = \frac{\tilde M}{\sqrt{|h|}} \buildrel{ |h| \to \infty }\over= 0 \quad ,  \quad
m = \frac{M^2}{M_{Pl}} = \frac{ {\tilde M}^2}{|h| \; M_{Pl}} \buildrel{ |h| \to \infty 
}\over= 0 \quad .
$$

\subsection{Regions of $n_s$ and $r$ covered by New Inflation and by 
Chaotic Inflation.}
\label{regions}
We see from eqs.(\ref{rtrino}), (\ref{cotrino}) and  (\ref{nsrtrikG}) that new inflation 
for $ h \leq 0 $ covers the narrow {\bf banana-shaped} sector between the black lines in 
the $(n_s,r)$ plane depicted in fig. \ref{ene}. We have in this region:
\be\label{cotsup}
0 < r < \frac8{N}  \quad , \quad n_s < 1 - \frac{1.9236 \ldots}{N} 
\quad {\rm for}\quad  \infty > y > 0 \; .
\ee
Chaotic inflation in the $ h < - 1 $ region covers the the even narrower
{\bf complementary} strip for $ \frac8{N} < r < \frac{16}{N} $
eq.(\ref{regir}).  The zero coupling point $ y \to 0 \; , \; r = \frac8{N}
\; , \; n_s = 1 -\frac2{N} $ being the border between the two regimes.

\medskip

Chaotic inflation for $ -1 < h \leq 0 $ covers a wide region depicted by the red lines
in the  $(n_s,r)$ plane [see fig. \ref{ene}]. However, as shown by fig. \ref{pdfzh},
this wide area is only a small corner of the field $ z $ - asymmetry $ h $ plane.
Since new inflation covers the banana-shaped region between the black curves, we see from
this figure that the most probable values of $ r $ are {\bf definitely non-zero} within 
trinomial new inflation. Precise lower bounds for $ r $ are derived from MCMC in eq.(\ref{cotinfr}).

\section{Monte Carlo Markov Chains and data analysis with the trinomial inflation models}

We performed Monte Carlo Markov Chains (MCMC) analysis of the commonly
available CMB and LSS data using the CosmoMC program \cite{mcmc}. For CMB
we considered the three years WMAP data, which provide the dominating
contribution, and also small scale data (ACBAR, CBI2, BOOMERANG03). For LSS
we considered SDSS (DR4). We used the second release of WMAP likelihood code.
In all our MCMC runs we did not marginalize over the SZ amplitude and did not
include non-linear effects in the evolution of the matter spectrum. 
The relative corrections are in any case not significant \cite{WMAP3},
especially in the present context.

\medskip

Given a theoretical model with several free parameters and a
sufficiently rich set of empirical data, MCMC is a very efficient stochastic numerical 
method to approximately reconstruct the probability distribution for
the actual values of those parameters. It is especially useful when the
theoretical predictions are themselves stochastic in nature, as is the case
of the primordial fluctuations that seed the CMB anisotropies.  CosmoMC is a
publicly available FORTRAN program that performs MCMC analysis over the
parameter space of the Standard Cosmological model and variations thereof.
The spectral index $ n_s $ of the adiabatic fluctuations and the ratio $r$ of
tensor to scalar fluctuations are two such parameters.  In CosmoMC usage,
one usually restrict the search to a subset of the full parameter space by
imposing so--called hard constraints, whose number and type depend on
previously acquired information both experimental and theoretical.

\medskip 

Concerning statistical convergence tests, our MCMC data were collected in parallel 
Message Passing Interface (MPI) runs with 10 to 16
chains, with the `R-1' stopping criterion (which looks at the fluctuations among parallel 
chains to decide when to stop the run) set to 0.03. Typically, this
meant single chains with length of the order $ 10^5 $, with the Gelman--Rubin
convergence test [var(chain mean)/mean(chain var) in the second half of the
chains] ranging from $ 10^{-3} $ to $ 10^{-2} $ for the various MCMC
parameters.

\medskip 

We imposed as a hard constraint that $ r $ and $ n_s $ are given by the
analytic expressions at order $ 1/N $ for the trinomial potential of the
inflaton. Namely, $ r $ and $ n_s $ are given by
eqs.(\ref{nscao})-(\ref{rcao}) for chaotic inflation and by
eqs.(\ref{nstrino})-(\ref{rnue}) for new inflation.
Our analysis differs in this {\bf crucial} aspect from previous MCMC studies
involving the WMAP3 data \cite{otros}. As natural within inflation, we also
included the inflationary consistency relation $ n_T = -r/8 $ on the tensor
spectral index. This constraint is in any case practically negligible.

\medskip

We allowed seven cosmological parameters to vary in our MCMC runs:
the baryonic matter fraction $ \omega_b $, the dark matter fraction
$ \omega_c $, the optical depth $ \tau $, the ratio of the (approximate) sound
horizon to the angular diameter distance $ \theta $, the primordial
superhorizon power in the curvature perturbation at $ 0.05~{\rm Mpc}^{-1} $,
$ A_s $, the scalar spectral index $ n_s $ and the tensor-scalar ratio $ r $.

For comparison, we also run MCMC's within the standard $\Lambda$CDM model
augmented by the tensor-scalar ratio $ r $ ($ \Lambda$CDM+$r $ model).  That
is, we treated $ n_s $ and $ r $ as unconstrained Monte Carlo parameters,
using standard priors. This analysis is indeed by now quite standard and
good priors are available already in CosmoMC. 

\medskip

We allowed the same seven parameters to vary in the MCMC runs for chaotic
and new inflation. 

\medskip

In the case of new inflation, since the characteristic banana--shaped allowed
region in the  $(n_s,r)$ plane is quite narrow and non--trivial, we used the two independent
variables $ z $ and $ h $ in trinomial inflationary setup as MC parameters.
That is, we used the analytic expressions at order $ 1/N $, eqs.(\ref{nstrino}) and (\ref{rnue})
to express  $ n_s $ and $ r $ in terms of  $ z $ and $ h $. To be more precise,
rather than $ z $ we used the appropriate normalized variable
\begin{equation} \label{eq:z1}
  z_1 = 1 - \frac{z}{z_+} = 1 - \frac{z}{\big(\sqrt{h^2 + 1} + |h|\big)^2} 
\end{equation}
$ z $ contains the field at horizon crossing and the coupling $ y $.
$ z_+ $ stands for $ z $ at the absolute minimum of the potential.
The variable $ z_1 $
grows monotonically from $0$ to $1$ as the coupling $y$ grows from
$0$ to $\infty$ [see eq.~(\ref{ntrino}) and the lines right below it].

Concerning priors, we kept the same, standard ones, of the $\Lambda$CDM+$r$
model for the first five parameters ($ \omega_b, \; \omega_c, \; \tau, \;
\theta $ and $ A_s $), while we considered all the possibilities for $ z_1 $
and $ h $, that is $ 0<z_1<1 , \; 0<h<\infty $.

\medskip

In the case of chaotic inflation we kept $ n_s $ and $ r $ as MC
parameters, imposing as hard priors that they lay in the region described by 
chaotic inflation [see fig.~\ref{ene}]. This is technically convenient, since this region
covers the major part of the probability support of $ n_s $ and $ r $ in
the $\Lambda$CDM+$r$ model and the parametrization 
eqs.(\ref{nscao})-(\ref{rcao}) in terms of the coupling parameters 
$ z $ and $ h $ becomes quite singular in the limit $ h \to -1 $. This is
indeed  the limit which allows to cover the region of highest likelihood.  

The distributions for the field variable $ z $ and the asymmetry parameter
$ h $ were obtained from the $ (n_s, \; r) $ distributions by a numerical
change of variables using eqs.(\ref{nscao}) and (\ref{rcao}) and taking
care of the corresponding Jacobian. We used a large collection of parallel
chains with a total number of samples close to five million.  In fact we
did not need to explicitly compute the Jacobian.  It was taken into account
automatically since we used an uniform two dimensional grid in the $z-h$
plane. Quite naturally in this approach for chaotic inflation, the most
likely values of the cosmological parameters and the corresponding maximum
of the likelihood coincide to those of the $ \Lambda$CDM+$r $ model.
 
The priors on the other parameters where the same of the $ \Lambda$CDM+$r$ 
model and of new inflation.

\medskip 

In all our MCMC runs we keep fixed the number of efolds $ N $
since horizon exit till the end of inflation. The reason is that the main
physics that determines the value of $ N $ is {\bf not} contained in the
available data but involves the reheating era. Therefore, it is {\bf not}
reliable to fit $ N $ solely to the CMB and LSS data.

The precise value of $ N $ is certainly near $ N =50 $ \cite{libros2,Ne}.
We take here the value $ N =50 $ as a reference
baseline value for numerical analysis, but from the explicit expressions
obtained in the slow roll $ 1/N $ expansion  
we see that both $ n_s - 1 $ and $ r $ scale as $ 1/N $.

Therefore, decreasing or increasing $ N $
produces a scale trasformation in the $(n_r-1,r)$ plane, thus displacing
the black and red curves in fig.  \ref{ene} towards up and left or towards down and
right.  This produces however small quantitative changes in our bounds for $
r $ as well as in the most probable values for $ r $ and $ n_s $. 
We ran MCMC simulations with variable $N$ and imposing
the trinomial new inflation potential. As a result, we found that $ N \sim
50 $ was the most probable value. A sharp fall in the likelihood happens
for $ N < 50 $ and a significant decrease shows up for $ N > 50 $. 
In summary, varying $ N $ has no dramatic effects and $ N \sim 50 $ 
turns to be the most probable value. 
The detailed  analysis with variable $ N $ is beyond the scope of the present paper.

\medskip 

We have not introduced the running of the spectral index $ dn_s / d\ln k $
in our MCMC fits since the running [eq. (\ref{indi})] must be very small of
the order $ {\cal O}(N^{-2}) \sim 0.001 $ in slow-roll and for generic
potentials \cite{1sN}. Indeed we found that adding $ dn_s / d\ln k $, as
given by eq.~(\ref{runcao}) or eq.~(\ref{runnew}), to the MCMC analysis
yields insignificant changes on the fit of $ n_s $ and $ r $.

On the contrary, when the running is introduced as a free parameter, then the fit
of $ n_s $ and $ r $ gets worse and values for the running much larger than
$ {\cal O}(N^{-2}) \sim 0.001 $ follow \cite{WMAP3,otros}.  We think that
the present data are {\bf not} yet precise enough to allow a determination
of $ dn_s / d\ln k $.  That is, adding further parameters to the fit (like
the running) does not improve the fit and does not teach anything new.

\subsection{MCMC results for new inflation.}

\begin{figure}[ht]
\includegraphics[width=14cm]{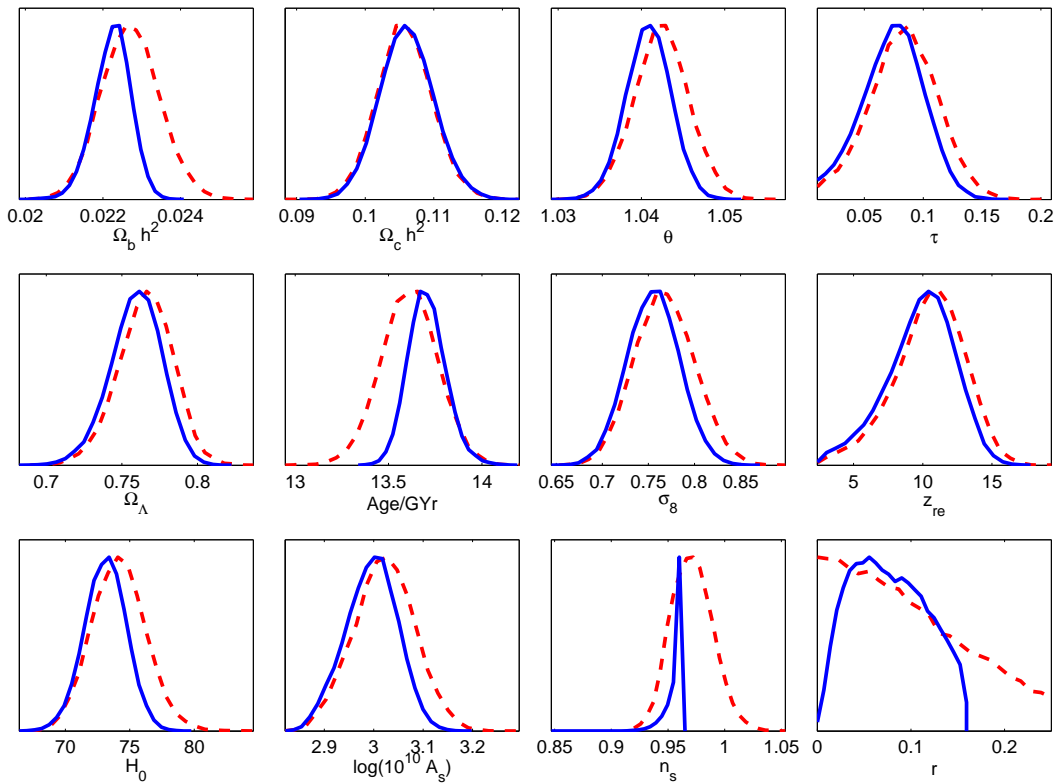}
\caption{Comparison of the marginalized probability distributions
  (normalized to have maximum equal to one) of the most relevant
  cosmological parameters (both primary and derived) between Trinomial New
  Inflation (solid blue curves) and the $ \Lambda$CDM+$r $ model (dashed red
  curves).  }
\label{new_vs_std}
\end{figure}

Our results for the new inflation trinomial potential eq.(\ref{trino})
are summarized in figs.~\ref{new_vs_std} and \ref{z1yhnsr}. 

\begin{figure}[ht]
\includegraphics[width=14cm]{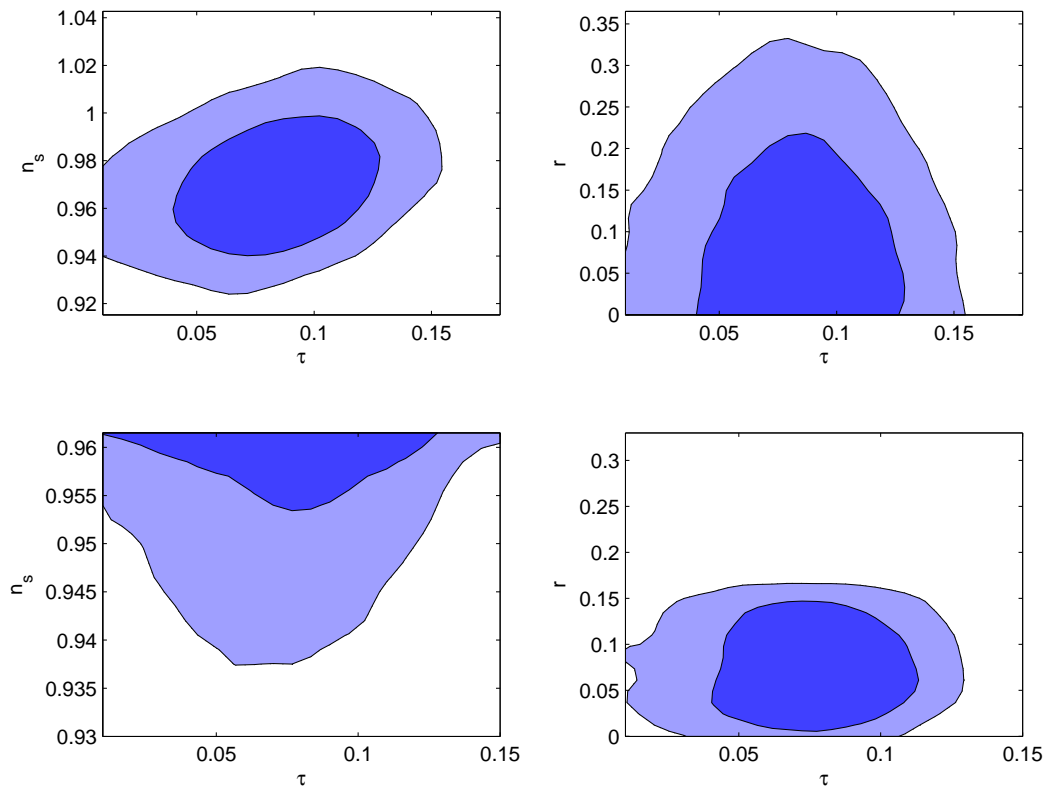}
\caption{Upper panels: 95\% and 68\% percent contour plots of joint
  probability $(\tau,n_s)$ distribution (left) and $(\tau,r)$ distribution
  (right) in the $\Lambda$CDM+$r$ model. Lower panels: the two joint
  distributions in trinomial new inflation. Recall that in this case $ n_s <
  0.9615\ldots $ is a theoretical bound when $N=50$.}
\label{taunsr}
\end{figure}

\begin{figure}[ht]
\includegraphics[width=14cm]{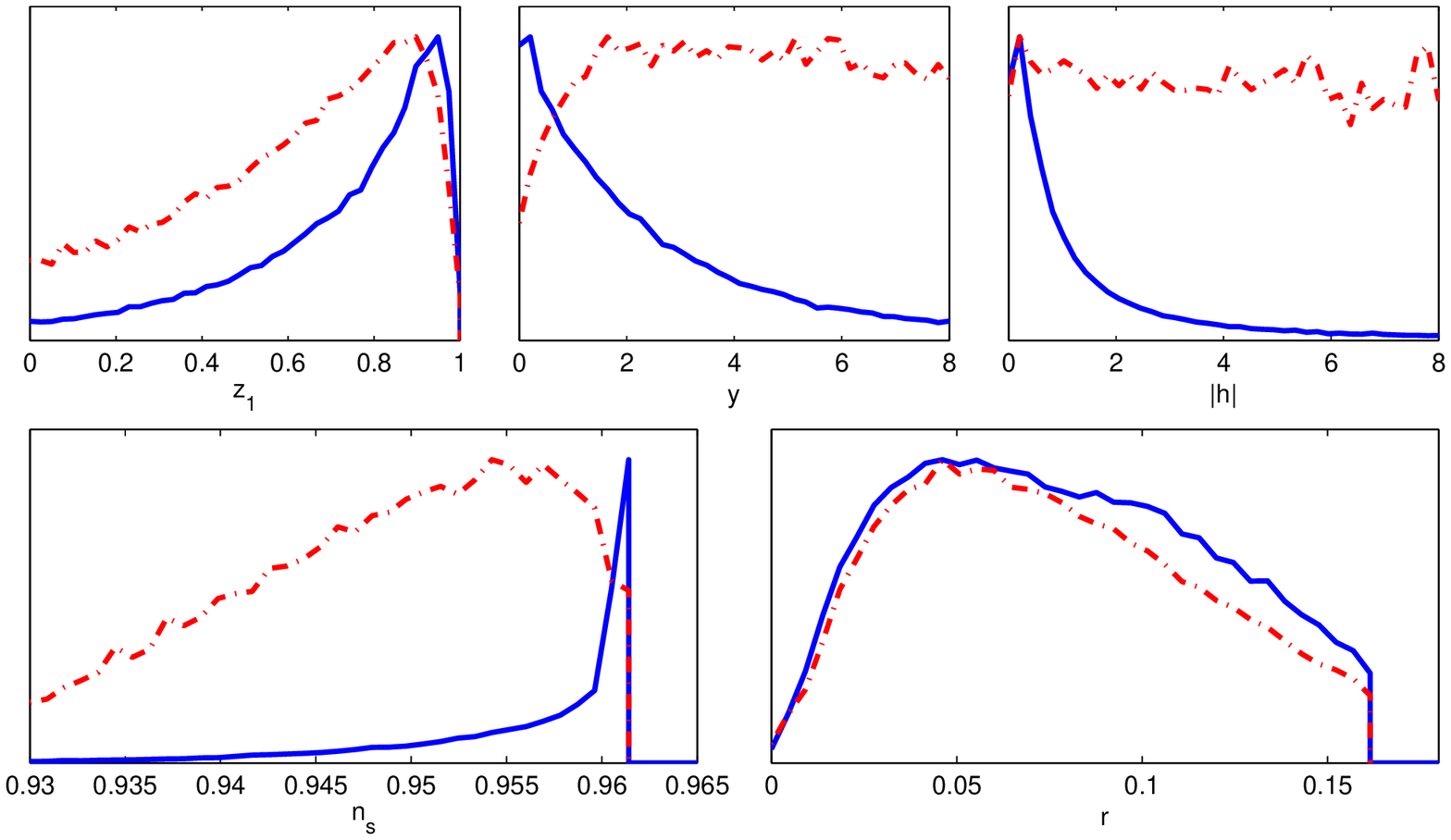}
\caption{Trinomial New Inflation. Upper panels: probability distributions
  (solid blue curves) and mean likelihoods (dot-dashed red curves), all
  normalized to have maximum equal to one, for the values of the normalized
  coupling at horizon exit $ z_1 $, of the quartic coupling $ y $ and of the
  modulus $ |h| $ of the asymmetry of the potential. Lower panels:
  probabilities and mean likelihoods for the values of $ n_s $ and $ r $.
  Notice that here $ n_s < 0.9615\ldots $ and $ r \le 0.16 $ since these
  are the theoretical upper bounds for $ n_s $ and $ r $ in trinomial new
  inflation with $ N=50 $. Trinomial new inflation exhibits a {\bf lower
    bound} for the ratio: $ r > 0.016 \; (95\% \; {\rm CL}) ,\; r > 0.049
  \; (68\% \; {\rm CL}) $. }
\label{z1yhnsr}
\end{figure}

In fig.~\ref{new_vs_std} we plot the marginalized probability distributions
(normalized to have maximum equal to one) of the most relevant cosmological
parameters, which are the primary ones allowed to vary independently in the
MCMC runs and some derived ones. The solid blue curves refer to the runs
with the hard priors specific of trinomial new inflation.  The dashed red
curves are those of the $ \Lambda$CDM+$r $ model. As should have been
expected from fig.~\ref{ene}, the really significant changes are restricted
solely to $ n_s $ and $ r $. To provide further evidence of this, we
compare in fig. ~\ref{taunsr} the joint probability
$ (\tau,n_s) $ and $ (\tau,r) $ distributions of trinomial new inflation with
those of the $ \Lambda$CDM+$r $ model. We recall that the optical depth parameter
$\tau$ is strongly correlated with $ n_s $.

In the lower panels of fig.~\ref{z1yhnsr} we provide an enlarged version of
the marginalized probability distributions for $ n_s $ and $ r $, together with
their mean likelihoods. 

Notice the difference between probability
distributions and mean likelihood distributions in the first four panels of
both figs. \ref{z1yhnsr} and \ref{zyhnsr}. Probability and
mean likelihood depend quite differently on the shapes and parametrizations
of the allowed regions; if observational data were well concentrated within
the allowed regions, both types of distributions would have a gaussian-like
shape and the difference would be much smaller. With the current data, the
joint probability distributions over the parameters of the trinomial
potential are very far from gaussian, as can be appreciated from fig. \ref{pdfzh}.

The theoretical constraints narrow the allowed region of parameters in a
very nontrivial way. Otherwise, the parameters could cover a much wider
region. Hence, the theoretically constrained
distributions can hardly be and in fact are not gaussian.
In the case of trinomial new inflation, the narrow banana--shaped region depicted
in fig. \ref{ene} is responsible for the marked difference 
between marginalized probability distribution and mean likelihood
for  $ n_s $, as shown in fig. \ref{z1yhnsr}.
Besides, the banana--shape of trinomial new inflation in the $ n_s-r $ plane,
produces a spike in the left lower panel of fig. \ref{z1yhnsr}. The sharp cut on the right is due 
to the theoretical upper bound on $ n_s $ given by eq. (\ref{cotsup}), the sharp rise on the left
of the maximum is due to the marginalization over $ r $.

We find a {\bf lower bound} on $ r $ for the trinomial new inflation model:
\be \label{cotinfr}
r > 0.016 \quad (95\% \;   {\rm CL}) \quad ,\quad r > 0.049 \quad 
(68\% \;  {\rm CL}) \quad {\rm (new \; inflation} )\; .
\ee
For $ n_s $ we find:
$$
n_s > 0.945 \quad (95\%  \;  {\rm CL}) \quad {\rm (new \; inflation)} \; .
$$
The most likely values, which can be read off directly from MCMC
data, are
$$
n_s \simeq 0.956 \quad ,\quad r\simeq 0.055 \quad {\rm (new \; inflation}) \; .
$$ 
The results for max(-log(likelihood)) are as follows:

\medskip

$\Lambda$CDM$+r$ model : 2714.038

\medskip

Trinomial chaotic inflation : 2714.038 $ \qquad $ and  $ \qquad $ Trinomial new inflation: 2714.153 \\

At the point with  most likely values for the cosmological parameters we
see a small increase of $ 0.115 $ of $ -\log$(likelihood$) $ for new inflation
with respect to the $\Lambda$CDM+$r$ model.
This increase is only twice as much of the standard deviation of 
max($-\log$(likelihood)) over ten parallel chains, which is 0.062 in the
case of the $\Lambda$CDM+$r$ model. Therefore, this increase of the 
$ -\log$(likelihood$) $ for new inflation is not significant.

\medskip

We recall that for trinomial new inflation there exist the theoretical
upper limits: $ n_s \leq 0.9615\ldots , \; r \leq 0.16$ \cite{nos2} .  That
is, the {\em most probable} value of $ n_s $ for trinomial new inflation is
very close to its {\em theoretical limiting} value and that of $ r $ is not
too far from it (see also fig.~\ref{z1yhnsr}).

\medskip 

In the upper panels of fig.~\ref{z1yhnsr} we plot the marginalized
probabilities and the mean likelihoods for the normalized coupling at
horizon exit $ z_1 $ and for $ y $ and $ h $, that is the two parameters of the
trinomial potential of new inflation. Their most likely values in our MCMC
runs, 
$$
z_1 \simeq 0.886, \quad y \simeq 2.01 \quad  {\rm and} \quad  
h \simeq 0.266 \quad {\rm (new \; inflation}) \; , 
$$
are not extremely significant  in view of the shapes of probabilities and mean likelihoods in
fig.~\ref{z1yhnsr}. More appropriate quantifiers are the upper bounds
\begin{equation}
  y < 2.46 \;,\quad h < 1.2  \quad {\rm with} \; 68\% \;  {\rm CL} \quad , \quad
  y < 6.35 \;,\quad h < 4.92 \quad {\rm with} \; 95\% \;  {\rm CL}\quad {\rm (new \; inflation})
\end{equation}
The $ z_1 $ distribution decreases for small $ z_1 $ (that is, where one approaches
the quadratic monomial potential at  $ y = 0 $) because such case yields a too large value 
for $ r , \;  (r \to 0.16) $ with respect to the experimental data.  
For $ \tilde z \to 1 $ the $ \tilde z $ distribution
decreases too because in this case one gets a too small spectral index $ n_s $.

\medskip 
In conclusion, the most likely trinomial potentials for new inflation are
almost symmetric and have moderate nonlinearity with the quartic coupling
$y$ of order $1$. The $ \chi \to - \chi $ symmetry is here broken since the
absolute minimum of the potential is at $ \chi \neq 0 $.

\begin{figure}[ht]
\includegraphics[width=14cm]{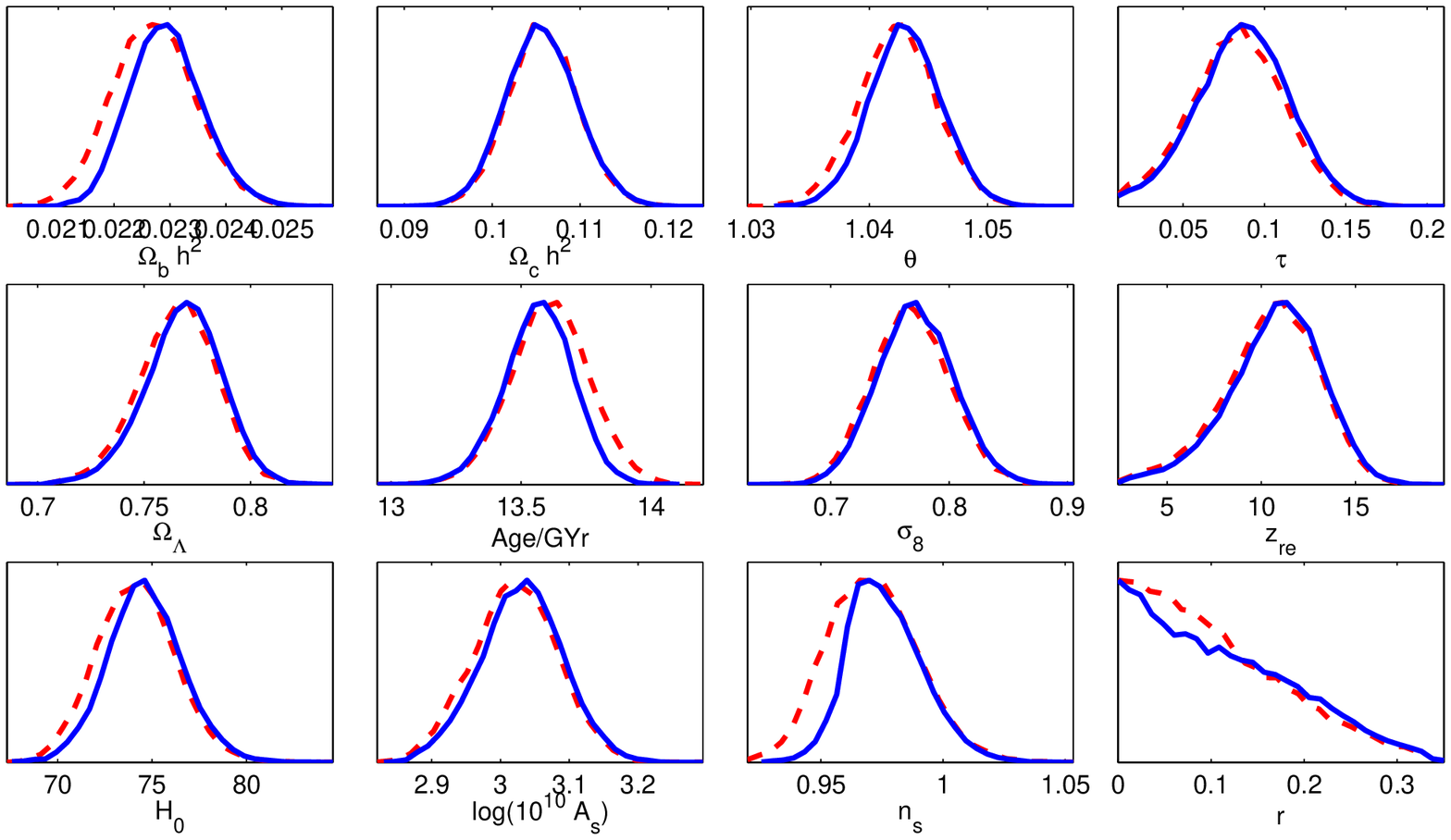}
\caption{Comparison of the marginalized probability distributions
  (normalized to have maximum equal to one) of the most relevant
  cosmological parameters (both primary and derived) between Trinomial Chaotic
  Inflation (solid blue curves) and the $ \Lambda$CDM+$r $ model (dashed red
  curves).  }
\label{chao_vs_std}
\end{figure}

\begin{figure}[ht]
\includegraphics[width=14cm]{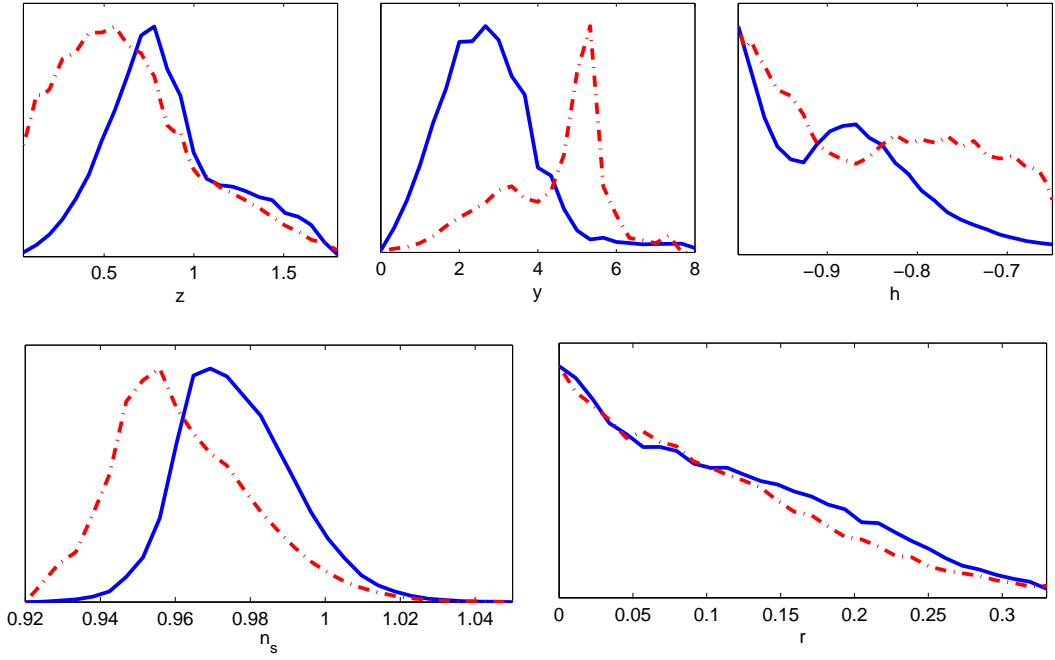}
\caption{Trinomial Chaotic Inflation. Upper panels: probability distributions
  (solid blue curves) and mean likelihoods (dot-dashed red curves), all
  normalized to have maximum equal to one, for the values of the normalized
  coupling at horizon exit $ z $, of the quartic coupling $ y $ and of the
  asymmetry $ h $ of the potential. Lower panels: probabilities and mean
  likelihoods for the values of $ n_s $ and $ r $. The data request a
  strongly asymmetric potential in chaotic inflation. That is, a strong
  breakdown of the $ \chi \to - \chi $ symmetry. }
\label{zyhnsr}
\end{figure}

\begin{figure}[ht]
\includegraphics{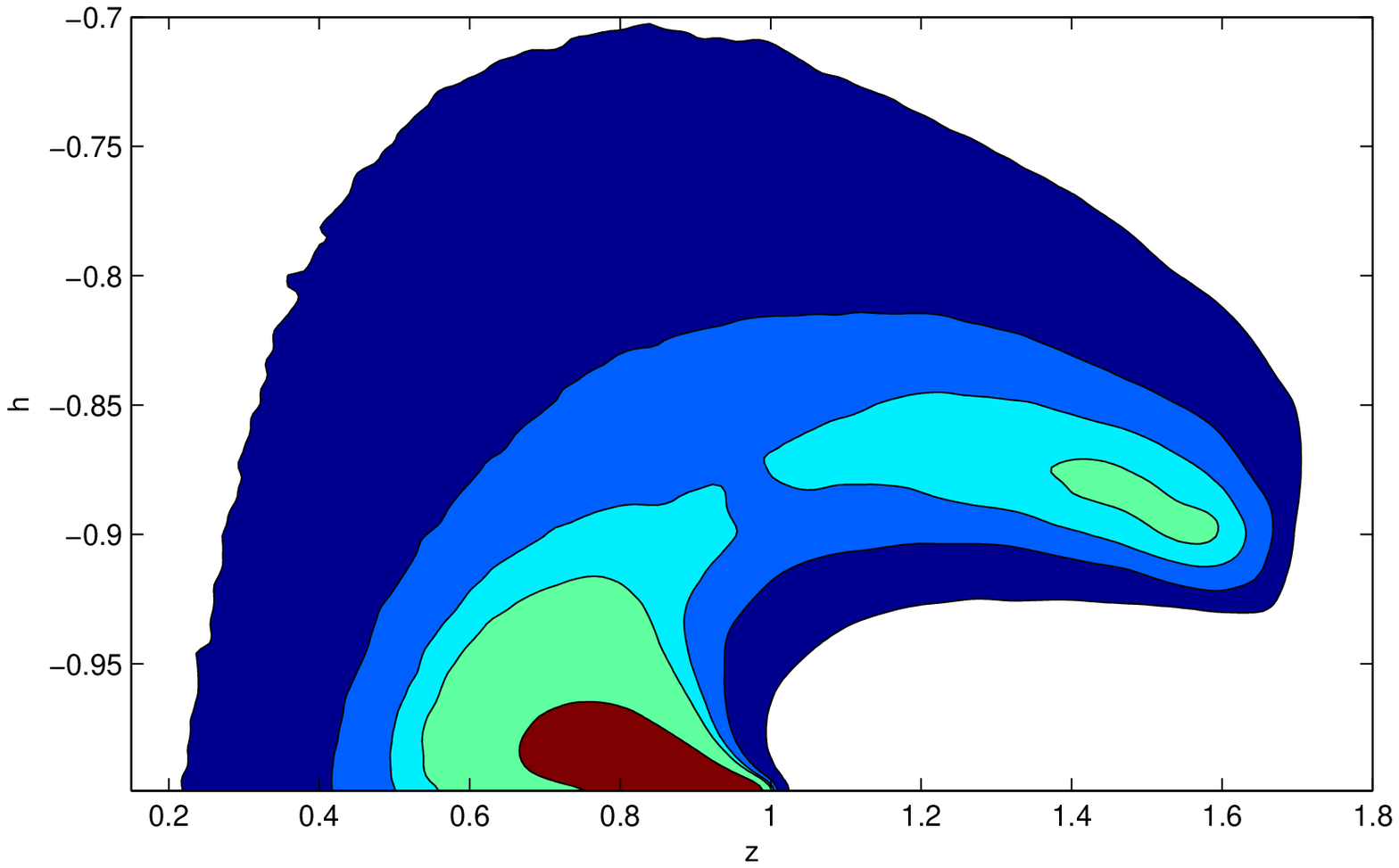}
\caption{Trinomial Chaotic Inflation.  12\%, 27\%, 45\%, 68\% and 95\%
confidence levels of the probability distribution in the field $ z $ - asymmetry $ h $ plane.
The color of the areas goes from brown to dark blue for increasing CL.
We see a strong preference for a very asymmetric potential with $ h < -0.95 $
and  a significative nonlinearity $ 0.7 < z < 1 $ in chaotic inflation.}
\label{pdfzh}
\end{figure}

\subsection{MCMC results for chaotic inflation.}

Our results for the trinomial potential of chaotic inflation, eq.(\ref{trino}),
are summarized in figs.~\ref{chao_vs_std}, \ref{zyhnsr} and \ref{pdfzh}. 

In fig.~\ref{chao_vs_std} we plot the marginalized probability
distributions (normalized to have maximum equal to one) of the usual
cosmological parameters as in fig.~\ref{new_vs_std}. Again, solid blue
curves refer to the runs with the hard priors of chaotic inflation, while
the dashed red curves are those of the $\Lambda$CDM+$r$ model. As should
have been expected from fig.~\ref{ene}, there are no really significant
changes in any parameter.

In the upper panels of fig.~\ref{zyhnsr} we plot the marginalized
probabilities and the mean likelihoods for the parameters of the trinomial
potential of chaotic inflation, the normalized coupling at horizon exit
$z$, the quartic coupling $y$ and the asymmetry $h$. These are numerically
calculated from those of $n_s$ and $r$ (which are also reported in the
lower panels of fig.~\ref{zyhnsr}), by means of eqs. (\ref{nscao}), (\ref{rcao}) 
and (\ref{2trino}). The strong non--linearities of these
relations are responsible for the peculiar shapes of the probabilities and
mean likelihoods, and their marked relative differences, of the variables
in the upper panels.

Finally, in fig. \ref{pdfzh} we depict the confidence levels at $ 12\%, \;
27\%,\; 45\%,\; 68 \%$ and $ 95 \% $ in the $ (z , \; h) $ plane. The
chaotic symmetric trinomial potential $ h = 0 $ is almost certainly {\bf
  ruled out} since $ h < -0.7 $ at 95 \% confidence level.

We see that the maximun probability is for strong asymmetry $ h \lesssim
-0.9 $ and significant nonlinearity $ 0.5 \lesssim z \lesssim 1 $ and $ 2.5
\lesssim y \lesssim 5$. That is, {\em all three} terms in the trinomial
potential $ w(\chi) $ {\em do contribute to the same order}.

Notice that the range $ 0.5 < z < 1 $ corresponds for $ h \to -1^+ $ to the
region where the quartic coupling is quite significant: $ 4.207\ldots < y <
+\infty $ according to eq.(\ref{yL}).

The probability is maximum on a highly special and narrow corner in the $(z,h)$ 
parameter space. This suggests: 
\begin{itemize}
\item[(i)] the data force both the asymmetry as well as the coupling to be {\bf large}.
\item[(ii)] the fact that $ z \sim 1 $ implies large $ y $ means that we
  are in the nonlinear regime in $ z $, that is strong coupling in $ \chi
  $. This suggests that higher order terms may be added and will be
  relevant.
\item[(iii)] If the preferred values are near the boundary of the parameter
  space, it could be that the true potential is {\bf beyond} that boundary.
  The region of parameter space in which trinomial chaotic inflation yields
  $ r \ll 1 $ is very narrow and highly non-generic and corresponds to an
  inflaton potential contrary to the Landau-Ginsburg spirit since adding
  higher degree terms in the field to the trinomial potential can produce
  relevant changes.  This means that the fit for chaotic inflation can be
  unstable while for new inflation the maximun probability happens for a
  moderate nonlinearity and therefore will not be much affected by higher
  degree monomials. Finally, the $ r \to 0 $ regime is obtained near the
  singular point $ z = 1, \; h = - 1 $ where the inflaton potential
  vanishes. 
\item[(iv)]The MCMC runs appear to go towards the $ h = - 1 $ limiting
  chaotic potential exhibiting an inflexion point [see fig. \ref{vv}] which
  is at the boundary of the space of parameters. This may indicate that the
  true potential is {\bf not} within the class of chaotic potentials. Since
  the runs go towards the maximal value of the asymmetry parameter $ h $
  where the inflaton potential exhibits an inflexion point, this supports
  the idea that the true potential {\bf must} definitely break the $ \chi
  \to - \chi $ symmetry, as new inflaton potentials do spontaneously.  This
  favours again new inflation since the best fit to the trinomial new
  inflation potential corresponds to small or zero asymmetry parameter $ h
  $. The spontaneous breakdown of the symmetry seems enough to obtain the
  best fit to the data without the presence of an explicit symmetry
  breaking term $ \frac{h}3 \; \sqrt{\frac{y}2} \; \chi^3 $ [see
  eq.(\ref{trino})].
\end{itemize}

\subsection{Conclusions.}

The MCMC results for chaotic and new inflation presented above show that:
\begin{itemize}
\item[(i)] symmetry breaking is begged by the data.  Namely, in chaotic
  inflation the data ask for a strongly asymmetric potential since we find
  the maximun probability when $ h $ is near the extreme value $ h = - 1$.
  In chaotic inflation, the $ h = 0 $ case is almost certainly ruled out as
  noticed above.  In new inflation the symmetry is spontaneously broken by
  construction since the absolute minima is at $ \chi \neq 0 $.  Hence, in
  one way or the other the data {\bf request} a breakdown either explicit
  or spontaneous of the $ \chi \to - \chi $ symmetry.

  New inflation naturally satisfies this requirement since the spontaneous
  symmetry breaking can alone reproduce the data with a moderate
  nonlinearity.
\item[(ii)] Adding higher order terms to the trinomial new inflation
  potential will not affect the trinomial fit since it corresponds to a
  moderate nonlinearity $ z_1 = 0.886 $. On the contrary,
  adding higher order terms in trinomial chaotic inflation can change the
  trinomial fit since we find a significative nonlinearity $ 0.5 < z < 1 $
  with the trinomial chaotic potential.  Therefore, in chaotic inflation,
  each time a new term of higher degree is added to the potential, it
  changes significatively all the coefficients in the potential. This is
  against the Landau-Ginsburg spirit which is the framework of the
  effective theory of inflation.]
\item[(iii)] Inflaton potentials of degree higher than four have been
  recently investigated in ref. \cite{nos3}. It is shown in ref.
  \cite{nos3} that binomial potentials with quadratic and $ \chi^{2 \, n} $
  terms reproduce the data in domains which are larger for $ n=2 $ than for
  $ n > 2 $. In addition, for new inflation it is shown that the crucial
  region to reproduce the data is the neighbourhood of the absolute minima
  of the potential. That is, only the quadratic and somehow the quartic
  terms around $ \chi_+ $ are relevant to fit the data.

  \medskip
  In summary, the MCMC analysis of the CMB+LSS data
  indicates that new trinomial inflation is the best model.

  \medskip
  Trinomial new inflation gives a very good fit to the data (CMB+LSS) with
  a moderate nonlinearity and small or zero asymmetry parameter. That is,
  the potential which best fits the data is the spontaneously broken
  symmetry potential eq.(\ref{trino}) with $ h = 0 $ and $ y \simeq 2.01
  \ldots $
  $$
  w(\chi) = \frac{y}{32} \left(\chi^2 - \frac8{y}\right)^{\! 2} \; ,
  $$
  Moreover, we have for this potential, when $N=50$,
  $$
  n_s \simeq 0.956 \quad ,\quad r\simeq 0.055 \quad {\rm (new \; inflation)} \; .
  $$
\item[(iv)] More generally, we have a {\bf lower bound} on the ratio $ r $
  within the trinomial new inflationary potentials given by
  $$ 
  r > 0.016 \; (95\% \; {\rm CL}) ,\; r > 0.049 \; (68\% \; {\rm CL}) \quad
  {\rm (new ~ inflation)} \; .$$
\end{itemize}

\bigskip

{\bf Acknowledgment:} The Monte Carlo simulations were performed at the
Avogadro-Golgi clusters at CILEA and on the Atena cluster at Dipartimento di Fisica,
Universit\`a Milano-Bicocca.

\end{document}